\documentclass[fleqn,usenatbib]{mnras}

\usepackage[T1]{fontenc}
\usepackage{printlen}
\usepackage{float}
\usepackage{svg}
\usepackage{comment}
\usepackage{bm}

\DeclareRobustCommand{\VAN}[3]{#2}
\let\VANthebibliography\thebibliography
\def\thebibliography{\DeclareRobustCommand{\VAN}[3]{##3}\VANthebibliography}

\usepackage{graphicx}
\usepackage{amsmath}
\usepackage{amssymb}

\usepackage{caption}
\usepackage{graphicx}
\usepackage{booktabs}

\usepackage{dsfont}

\usepackage{newtxtext,newtxmath}

\title[Field-Level Inference with Emulators]{Bayesian Inference of Initial Conditions from Non-Linear Cosmic Structures using Field-Level Emulators}

\author[L. Doeser et al.]{
Ludvig Doeser$^{1}$\thanks{E-mail: ludvig.doeser@fysik.su.se},
Drew Jamieson$^{2}$, Stephen Stopyra$^{1}$, Guilhem Lavaux$^{3}$, Florent Leclercq$^{3}$, Jens Jasche$^{1,3}$ 
\\
$^{1}$The Oskar Klein Centre, Department of Physics, Stockholm University, Albanova University
Center, SE 106 91 Stockholm, Sweden\\
$^{2}$Max-Planck-Institut für Astrophysik, Karl-Schwarzschild-Straße 1, 85748 Garching, Germany\\
$^{3}$CNRS \& Sorbonne Université, UMR 7095, Institut d’Astrophysique de Paris, 98 bis boulevard Arago, F-75014 Paris,
France\\
}

\date{Accepted XXX. Received YYY; in original form ZZZ}

\pubyear{2023}

\begin{document}
\label{firstpage}
\pagerange{\pageref{firstpage}--\pageref{lastpage}}
\maketitle

\begin{abstract}
Analysing next-generation cosmological data requires balancing accurate modeling of non-linear gravitational structure formation and computational demands. We propose a solution by introducing a machine learning-based field-level emulator, within the Hamiltonian Monte Carlo-based Bayesian Origin Reconstruction from Galaxies (\texttt{BORG}) inference algorithm. Built on a V-net neural network architecture, the emulator enhances the predictions by first-order Lagrangian perturbation theory to be accurately aligned with full $N$-body simulations while significantly reducing evaluation time. We test its incorporation in \texttt{BORG} for sampling cosmic initial conditions using mock data based on non-linear large-scale structures from $N$-body simulations and Gaussian noise. The method efficiently and accurately explores the high-dimensional parameter space of initial conditions, fully extracting the cross-correlation information of the data field binned at a resolution of $1.95h^{-1}$ Mpc. Percent-level agreement with the ground truth in the power spectrum and bispectrum is achieved up to the Nyquist frequency $k_\mathrm{N} \approx 2.79h \; \mathrm{Mpc}^{-1}$. Posterior resimulations – using the inferred initial conditions for $N$-body simulations – show that the recovery of information in the initial conditions is sufficient to accurately reproduce halo properties. In particular, we show highly accurate $M_{200\mathrm{c}}$ halo mass function and stacked density profiles of haloes in different mass bins $[0.853,16]\times 10^{14}M_{\odot}h^{-1}$. As all available cross-correlation information is extracted, we acknowledge that limitations in recovering the initial conditions stem from the noise level and data grid resolution. This is promising as it underscores the significance of accurate non-linear modeling, indicating the potential for extracting additional information at smaller scales.
\end{abstract}


\begingroup
\let\clearpage\relax
\section{Introduction}

\begin{figure*}
    \centering
    \includegraphics{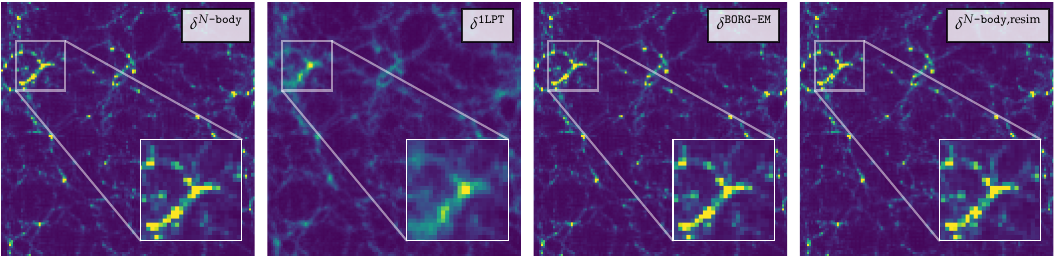}
    
    \caption{Visual comparison of three physical forward models at redshift $z=0$, illustrated in the panels as slices through the density field. In all cases, $128^3$ particles in a cubic volume with side length $250h^{-1}$ Mpc were simulated. Identical initial conditions underpin the $N$-body simulation, the first-order Lagrangian Perturbation Theory ($1$LPT), and the emulator (\texttt{BORG-EM}). The emulator, utilizing $1$LPT displacements as input and generating $N$-body-like displacements, effectively bridges the gap between $1$LPT and $N$-body outcomes; while the $1$LPT struggles to replicate collapsed overdensities as effectively as the $N$-body simulation, the emulator achieves remarkable success in reproducing $N$-body-like cosmic structures. In the right-most panel, the $N$-body simulation's initial conditions originate from the posterior distribution realized through the \texttt{BORG} algorithm. The inference process employs the $N$-body density field (the left-most panel) combined with Gaussian noise as the ground truth, utilizing the \texttt{BORG-EM} as the forward model during inference. As expected, given that this represents a single realization of posterior resimulations of initial conditions, minor deviations from the true $N$-body field are visible. Being able to capture large-scale structures as well as collapsed overdensities, the field-level emulator demonstrates that field-level inference from non-linear cosmic structures is feasible.}
\label{fig:emulator}
\end{figure*}

Imminent next-generation cosmological surveys, such as DESI \citep{DESICollaboration2016}, Euclid \citep{Laureijs2011,Amendola2018},
LSST at the Vera C. Rubin Observatory \citep{LSSTScienceCollaboration2009,LSSTDarkEnergyScienceCollaboration2012,Ivezic2019}, SPHEREx \citep{Dore2014}, and Subaru Prime Focus Spectrograph \citep{Takada2012}, will provide an unprecedented wealth of galaxy data probing the large-scale structure of the Universe. The increased volume of data, with the expected number of galaxies in the order of billions, must now be matched by accurate modelling to optimally extract physical information. Analysis at the field level recently emerged as a successful alternative 
to the traditional way of analysing cosmological data – through a limited set of summary statistics – and instead uses information from the entire field \citep{Jasche2012,Wang2014, Jasche2015, Lavaux2016}. 

Addressing the complexities of modelling higher-order statistics and defining optimal summary statistics for information extraction from the galaxy distribution is a challenge that can be overcome by employing a fully numerical approach at the field level \citep[see e.g.,][]{Jasche2015, Lavaux2016, Jasche2019, Lavaux2019}. All higher-order statistics of the cosmic matter distribution are generated through a physics model, which connects the early Universe with the large-scale structure of today. Enabled by the nearly Gaussian nature of the initial perturbations, the inference is pushed to explore the space of plausible initial conditions from which the present-day non-linear cosmic structures formed under gravitational collapse. Although accurate physics models are needed, the computational resources required for parameter inference prompt the use of approximate models. In this work, we propose a solution by introducing a physics model based on a machine-learning-trained field-level emulator as an extension to first-order Lagrangian Perturbation Theory ($1$LPT) to accurately align with full $N$-body simulations, resulting in higher accuracy and lower computational footprint during inference than currently available.

Field-level inferences, supported by a spectrum of methodologies and nuances \citep[e.g.,][]{Jasche2012,Kitaura2013,Wang2014,Ata2014,Ata2020,Seljak2017,Jasche2017,Schmidt2018,Bos_2019,Kitaura2019,villaescusanavarro2021robust,Modi2018,Modi2022,Hahn2022,Hahn2022a,Kostic2022,Bayer2023,Bayer2023b,Legin2023, Jindal2023, Stadler2023}, have become a prominent approach. Notably, analysis at the field level has recently been shown to best serve the goal of maximizing the information extraction \citep{Leclercq2021,Boruah2023}. It has also been shown that a significant amount of the cosmological signal is embedded in non-linear scales \citep[e.g.,][]{Ma2016, Seljak2017, Villaescusa-Navarro2020, Villaescusa-Navarro2021b}, which can be optimally extracted by field-level inference. $N$-body simulations are currently the most sophisticated numerical method to simulate the full non-linear structure formation of our Universe \citep[for a review, see][]{Vogelsberger2020}. Direct use of $N$-body simulations in field-level inference pipelines is nonetheless challenging, due to the high cost of model evaluations. To lower the required computational resources, all the while resolving non-linear scales, quasi $N$-body numerical schemes have been developed, e.g. \texttt{tCOLA} \citep{Tassev2013}, \texttt{FastPM} \citep{Feng2016}, \texttt{FlowPM} \citep{Modi2021}, \texttt{PMWD} \citep{Ll2022} and Hybrid Physical-Neural ODEs \citep{Lanzieri2022}. All these, however, involve significant trade-offs between speed and non-linear accuracy \citep[e.g.][]{Stopyra2023}.

As demonstrated with the Markov Chain Monte Carlo (MCMC) based Bayesian Origin Reconstruction from Galaxies (\texttt{BORG}) algorithm \citep{Jasche2012} for galaxy clustering \citep{Jasche2015,Lavaux2016,Jasche2019,Lavaux2019}, weak lensing \citep{Porqueres2021,Porqueres2021a,Porqueres2023}, velocity tracers \citep{Prideaux-Ghee2022a}, and Lyman-$\alpha$ forest \citep{Porqueres2019,Porqueres2020}, field-level inference with more than tens of millions of parameters has become feasible. These applications, in addition to other studies within the \texttt{BORG} framework \citep{Ramanah2018,Jasche2019,Nguyen2020a,Andrews2022,Tsaprazi2021,Tsaprazi2023}, rely on fast, approximate, and differentiable physical forward models, such as first and second-order Lagrangian Perturbation Theory ($1$LPT/$2$LPT), non-linear particle mesh models and \texttt{tCOLA}. The application of the \texttt{BORG} algorithm \citep{Jasche2015,Lavaux2016, Lavaux2019,Jasche2019} to trace the galaxy distribution of the nearby Universe from SDSS-II \citep{Abazajian2008}, BOSS/SDSS-III \citep{Dawson2012,Eisenstein2011}, and the 2M++ catalog \citep{Lavaux2011}, respectively, further laid the foundation for posterior resimulations of the local Universe with high-resolution $N$-body simulations \citep{Leclercq2015,Nguyen2020a,Desmond2021,Hutt2022,Mcalpine2022,Stiskalek2023}. With the success of this approach, the requirements increased. Posterior resimulations require that the inferred initial conditions when evolved to redshift $z=0$ reproduce accurate cluster masses and halo properties. To achieve this goal, \citet{Stopyra2023} showed that a minimal accuracy of the physics simulator during inference is required. This inevitably leads to a higher computational demand, which we address in this work by the use of the field-level emulator.

Specifically, we propose to use a machine learning replacement for $N$-body simulations in the inference. It takes the form of a field-level emulator, which is trained through deep learning to mimic the predictions of the complex $N$-body simulation, as displayed in Fig.~\ref{fig:emulator}. Such emulators have recently been used both to map cosmological initial conditions to the final density field \citep{Bernardini2020} and to map the output of a fast and less accurate simulation to the more complex output of an $N$-body simulation \citep{He2019,AlvesDeOliveira,Kaushal,Jamieson2022,Jamieson2022b}. While \cite{He2019} and \cite{AlvesDeOliveira} map $1$LPT to \texttt{FASTPM} and $N$-body respectively for a fixed cosmology, \cite{Jamieson2022b} made significant advancement over the previous works by both training over different cosmologies and mapping $1$LPT directly to $N$-body simulations. Similarly, \citet{Kaushal} presented the mapping from \textsc{tcola} \texttt{tCOLA} to $N$-body for various cosmologies. The main benefit of the field-level emulator over other gravity models is being fast and differentiable to evaluate, all the while achieving percent level accuracy at scales down to $k \sim 1h$ Mpc$^{-1}$ compared to $N$-body simulations. The computational cost to model the complexity is instead pushed into the network training of the emulator.

An investigation of the robustness of field-level emulators was made in \cite{Jamieson2022}. The emulator achieves percent-level accuracy deep into the non-linear regime and generalizes well, demonstrating that it has learned general physical principles and nonlinear mode couplings. This builds confidence in emulators being part of the physical forward model within cosmological inference. Other recent works have also demonstrated promise in inferring the initial conditions from non-linear dark matter densities with machine-learning based methods \citep{Legin2023,Jindal2023}. 

The manuscript is structured as follows. In Section~\ref{sec:2} we describe the merging of the $1$LPT forward model in \texttt{BORG} with a field-level emulator, which we call \texttt{BORG-EM}, to infer the initial conditions from non-linear dark matter densities, which we discuss along the introduction of a novel multi-scale likelihood in Section~\ref{sec:3}. We demonstrate the efficient exploration of the high-dimensional parameter space of initial conditions in Section~\ref{sec:4}. In Section~\ref{sec:5} we show that our method fully extracts all cross-correlation information from the data well into the noise-dominated regime up to the data grid resolution limit. We use the inferred initial conditions to run posterior resimulations and show in Section~\ref{sec:6} that the recovered information in the initial conditions is sufficient for accurate recovery of halo masses and density profiles. We summarize and conclude our findings in Section~\ref{sec:7}.
\section{Field-level Inference with Emulators}
\label{sec:2}
Inferring the initial conditions of the Universe from available data is enabled by the Bayesian Origin Reconstruction from Galaxies (\texttt{BORG}) algorithm. The modular structure of \texttt{BORG} facilitates replacement of the forward model, which translates initial conditions to the data space. To increase the accuracy and decrease the computational cost we integrate field-level emulators. 

\subsection{The \texttt{BORG} algorithm}
The \texttt{BORG} algorithm is a Bayesian hierarchical inference framework designed to analyze the three-dimensional cosmic matter distribution underlying observed galaxies in surveys \citep{Jasche2010a,Jasche2012,Jasche2015,Lavaux2016,Jasche2019,Lavaux2019}. More specifically, by incorporating physical models of gravitational structure formation and a data likelihood \texttt{BORG} turns the task of analysing the present-day cosmic structures into exploring the high-dimensional space of plausible initial conditions from which these observed structures formed. To explore this vast parameter space, \texttt{BORG} uses the Hamiltonian Monte Carlo (HMC) framework generating Markov Chain Monte Carlo (MCMC) samples that approximate the posterior distribution of initial conditions \citep{Jasche2010a, Jasche2012}. The benefit of the HMC is two-fold: 1) it utilizes conserved quantities such as the Hamiltonian to obtain a high Metropolis-Hastings acceptance rate, effectively reducing the rejected model evaluations, and 2) it efficiently uses model gradients to traverse the parameter space, reducing random walk behaviour through persistent motion \citep{Duane1987a,Neal2011ProbabilisticIU}.

Importantly, \texttt{BORG} solely relies on forward model evaluations and at no point requires the inversion of the flow of time in them, which generally is not possible due to inverse problems being ill-posed because of noisy and incomplete observational data \citep{Nusser92,Crocce2005}. \texttt{BORG} implements a fully differentiable forward model such that the adjoint gradient with respect to the initial conditions can be computed. Current physics models in \texttt{BORG} include first and second-order Lagrangian Perturbation Theory (\texttt{BORG-1LPT}/\texttt{BORG-2LPT}), non-linear Particle Mesh models \texttt{BORG-PM} with/without \texttt{tCOLA}. During inference \texttt{BORG} accounts for both systematic and stochastic observational uncertainties, such as survey characteristics, selection effects, and luminosity-dependent galaxy biases as well as unknown noise and foreground contaminations \citep{Jasche2012,Jasche2015,Lavaux2016,Jasche2017}. As \texttt{BORG} infers the initial conditions, it effectively also recovers the dynamic formation history of the large-scale structure as well as its evolved density and velocity fields.

The sampled initial conditions generated by \texttt{BORG} can be used for running posterior resimulations using more sophisticated structure formation models at high resolution \citep{Leclercq2015,Nguyen2020a,Desmond2021,Mcalpine2022}. In this work, we aim to improve the fidelity of inferences through the incorporation of novel machine-learning emulator techniques into the forward model, as required for accurate recovery of massive structures \citep{Nguyen2020,Stopyra2023}. 

\begin{figure*}
    \centering
    \includegraphics{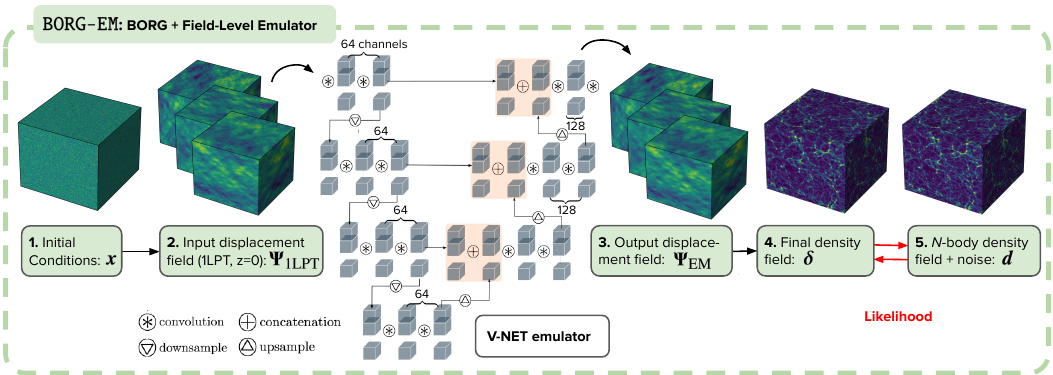}
    \caption{Overview of the field-level emulator integration within the \texttt{BORG} algorithm, which we call \texttt{BORG-EM}. The process begins with the evolution of the initial white-noise field using first-order Lagrangian Perturbation Theory ($1$LPT) up to redshift $z=0$, yielding the displacement field $\Psi_{1\mathrm{LPT}}$. After periodic padding, these displacements are passed through the V-NET emulator. The emulator's output displacement field $\Psi_{\mathrm{EM}}$ is subsequently transformed into an overdensity field $\boldsymbol{\delta}$ using a cloud-in-cell mass assignment scheme. A likelihood evaluation with the data $\mathbfit{d}$ (here the output of an $N$-body simulation $+$ Gaussian noise) ensues, followed by the back-propagation of the adjoint gradient for updating the initial conditions.}
    \label{fig:borg-em}
\end{figure*}

\subsection{Field-level emulators}
In this work, we build upon the work of \cite{Jamieson2022b} by incorporating the convolutional neural network (CNN) emulator into the physical forward model of \texttt{BORG}. While an $N$-body simulation aims to translate initial particle positions into their final positions, which represent the non-linear dark matter distribution, several approximate models strive for the same result with a lower computational cost. Specifically, our emulator aims to use as input first-order LPT and correct it such that the output of the emulator aligns with the results of an actual $N$-body simulation. 

Denoting $\mathbfit{q}$ as the initial positions on an equidistant Cartesian grid in Lagrangian space and $\mathbfit{p}$ as the final positions of simulated dark matter particles, we define the displacements as $\boldsymbol{\Psi} = \mathbfit{p}-\mathbfit{q}$. The $3$-dimensional displacement field $\boldsymbol{\Psi}_{1\mathrm{LPT}}$ generated by $1$LPT at redshift $z=0$ is used as input. Let us denote the emulator mapping $G_{\mathrm{EM}}$ as
\begin{equation}
    \boldsymbol{\Psi}_{\mathrm{EM}} = G_{\mathrm{EM}}(\boldsymbol{\Psi}_{1\mathrm{LPT}}),
\end{equation}
where $\boldsymbol{\Psi}_{\mathrm{EM}}$ is the output displacement field of the emulator. The final positions $\mathbfit{p}$ of the emulator are then given by 
\begin{equation}
\mathbfit{p}=\mathbfit{q}+\boldsymbol{\Psi}_{\mathrm{EM}}(\mathbfit{p}_{1\mathrm{LPT}},\mathbfit{q}).
    \label{eq:final_pos}
\end{equation} 
As we will see in section~\ref{sec:sub_model_retraining}, $\Omega_\mathrm{m}$ is an additional input to the network, which helps encode the dependence of clustering across different scales and enables the emulator to accurately reproduce the final particle positions (at redshift $z=0$) of $N$-body simulations for a wide range of cosmologies as shown in \citet{Jamieson2022b}. For this work, we use a fixed set of cosmological parameters and we will therefore not explicitly state $\Omega_\mathrm{m}$ as an input to the emulator.

\subsubsection{Model re-training}
\label{sec:sub_model_retraining}
The model architecture of the emulator is identical to the one described in \citet{Jamieson2022b} and is based on the U-Net/V-Net architecture \citep{Ronneberger,Milletari2016}. These specialized convolutional neural network architectures were initially developed for biomedical image segmentation and three-dimensional volumetric medical imaging, respectively. By working on multiple levels of resolution connected in a U-shape as shown in Fig~\ref{fig:borg-em}, first by several downsampling layers and then by the same amount of upsampling layers, these architectures excel at capturing fine spatial details. The sensitivity to information at different scales of these architectures is a critical aspect, making them particularly suitable for accurately representing cosmological large-scale structures. They also maintain intricate spatial information through skip connections in their down- and upsampling paths.  

Our neural network (whose architecture is described in more detail in Appendix \ref{app:NN}) is trained and tested using the framework \texttt{map2map}\footnote{\href{https://github.com/eelregit/map2map}{github.com/eelregit/map2map}} for field-to-field emulators, based on \texttt{PyTorch} \citep{Paszke2019}. Gradients of the loss function with respect to the model weights during training and of the data model with respect to the input during field-level inference are offered by the automatic differentiation engine \texttt{autograd} in \texttt{PyTorch}.

For detailed model consistency within the \texttt{BORG} framework, we retrain the model weights of the emulator using the \texttt{BORG-1LPT} predictions as input (for details see Appendix \ref{app:borglpt}). To predict the cosmological power spectrum of initial conditions, we use the \texttt{CLASS} transfer function \citep{Blas2011}. As outputs during training, we use the Quijote Latin Hypercube (Quijote LH) $N$-body suite \citep{Villaescusa-Navarro2020}, wherein each simulation is characterized by unique values for the five cosmological parameters: $\Omega_\mathrm{m},\Omega_\mathrm{b},\sigma_8,n_\mathrm{s},$ and $h$. As the emulator expects a $\Lambda$CDM cosmological background, it can navigate through this variety of cosmologies by explicitly using the $\Omega_\mathrm{m}$ as input to each layer of the CNN. The other parameters only affect the initial conditions, not gravitational clustering, and need not be included as input. In total, $2000$ input-output simulation pairs from \texttt{BORG-1LPT} and the Quijote latin-hypercube were generated, out of which $1757$, $122$, and $121$ cosmologies were used in the training, validation, and test set respectively. 

As described in \citet{Jamieson2022b}, even though the emulator was trained on $512^3$ particles in a $1h^{-1}$ Gpc box, the only requirement on the input is that the Lagrangian resolution is $1.95h^{-1}$ Mpc. The emulator can thus handle larger volumes by dividing them into smaller sub-volumes to be processed independently, with subsequent tiling of the sub-volumes to construct the complete field. For smaller volumes that can be handled directly, periodic padding of $48$ cells in each dimension is necessary before the network prediction. 

\subsubsection{Mini-emulator}
In this work, we also train and deploy a compact variant of the field-level emulator which we call mini-emulator, described in Appendix~\ref{app:distill}. The modified architecture of the emulator results in the number of model parameters being reduced by a factor of four. In turn, this reduces the forward and adjoint computations with a factor of four while not reducing the accuracy significantly (see Appendix \ref{app:fwd_models}). This makes the mini-emulator especially appealing for application during the initial phase of the \texttt{BORG} inference. This also suggests that improved network architectures can still yield improved performance. A detailed investigation of model architectures is, however, beyond the scope of this paper and will be explored in future works.

\subsubsection{Timing and accuracy of emulator}
We compare the improvements in timing and accuracy offered by the emulator against forward models previously used in \texttt{BORG}, such as \texttt{BORG-1LPT} and \texttt{COLA}\footnote{Further comparisons of interest with other fast GPU-based forward models \citep[e.g.,][]{Ll2022} would require a thorough evaluation of their timing and accuracy, as well as their trade-off, which is beyond the scope of this study.}. In Table~\ref{tab:time_comparison} we highlight the significant increase in computational efficiency achieved by our emulator \texttt{BORG-EM} and the mini-emulator.

\begin{table}
    \centering
    \begin{tabular}[width=1.0\linewidth]{@{}lll@{}}
    \toprule
    \textbf{Structure Formation Model} & $\mathbfit{t}_{\mathrm{forward}}$ [s] & $\mathbfit{t}_{\mathrm{adjoint}}$ [s] \\ 
    \midrule
        $1$LPT & $<10^{-1}$ & $<10^{-1}$ \\
        \texttt{BORG-EM} ($1$LPT $+$ emulator)$^\star$ & $1.6$ & $2.6$  \\
        $1$LPT $+$ mini-emulator$^\star$ & $0.4$ & $0.6$  \\
        \texttt{BORG-PM} (\texttt{COLA}, $n_{\mathrm{steps}} = 20$, forcesampling$=4$)$^{\dagger}$ & $\sim 8$ & $\sim 8$  \\
        $N$-body (\texttt{P-Gadget}-III, $n_{\mathrm{steps}} = 1664$)$^{\dagger}$ & $\sim 6 \times 10^2$ & – \\
    \midrule
    \midrule
\end{tabular}
    \caption{To demonstrate the effectiveness of the emulator, we compare evaluation times (in wall-clock seconds) of the forward and adjoint parts of different approximate physics models available in \texttt{BORG}, bench-marked against a full $N$-body simulation. The indicative times are of relevance for the generation of \texttt{BORG} samples of initial conditions per time unit. Comprehensive timing comparison requires optimization for specific settings and hardware in each method, which is outside the scope of this work. The scenario involves $128^3$ particles in a cubic volume of side length $250h^{-1}$ Mpc. $\star$: Supermicro 4124GS-TNR node with NVIDIA A100 40GiB, $\dagger$: $128$ cores on a Dell R6525 node with AMD EPYC Rome 7502 using the \texttt{COLA} settings required for sufficiently high accuracy as shown by \citet{Stopyra2023}.}
    \label{tab:time_comparison}
\end{table}

In terms of accuracy, as shown in \cite{Jamieson2022b}, the field-level emulator achieves percent level accuracy for the density power spectrum, the Lagrangian displacement power spectrum, the momentum power spectrum, and density bispectra as compared to the Quijote suite of $N$-body simulations down to $k\sim 1h$ Mpc$^{-1}$. In Appendix~\ref{app:powspec_and_bispec} we define some of these summary statistics \citep[also see][]{Jamieson2022b}. High accuracy is also achieved for the halo mass function, as well as for halo profiles and the matter density power spectrum in redshift space. 

In Fig.~\ref{fig:emulator}, we present a graphical comparison between the results obtained from \texttt{BORG-1LPT} with and without the utilization of the re-trained emulator \texttt{BORG-EM}, in addition to the corresponding $N$-body simulation utilizing \texttt{P-Gadget-III} (see more in section \ref{sec:groundtruth}). The emulator effectively collapses the overdensities of the \texttt{BORG-1LPT} prediction into more pronounced haloes, filaments, and walls, forming the cosmic web, while expanding underdense regions into cosmic voids. In Appendix \ref{app:fwd_models} we show the accuracies in terms of the cross-power spectrum, power spectrum, bispectrum, halo mass function, and stacked halo density profiles. 

\subsection{\texttt{BORG} + Field-level emulator}
In general, \texttt{BORG} obtains data-constrained realizations of a set of plausible three-dimensional initial conditions in the form of the white noise amplitudes $\mathbfit{x}$ given some data $\mathbfit{d}$, such as a dark matter over-density field or an observed galaxy counts. Following \citet{Jasche2019}, one can show that the posterior distribution from Bayes Law reads
\begin{equation}
    \pi\left(\mathbfit{x} | \mathbfit{d}\right)  = \frac{\pi\left(\mathbfit{x}\right) \pi\left(\mathbfit{d} | G(\mathbfit{x},\boldsymbol{\Omega})\right)}{\pi\left(\mathbfit{d}\right)},
\end{equation}
where $\pi\left(\mathbfit{x}\right)$ is the prior distribution encompassing our a priori knowledge about the initial white-noise field, $\pi\left(\mathbfit{d}\right)$ is the evidence which normalizes the posterior distribution, and $\pi\left(\mathbfit{d} | G(\mathbfit{x},\boldsymbol{\Omega})\right)$ is the likelihood that describes the statistical process of obtaining the data $\mathbfit{d}$ given the initial conditions $\mathbfit{x}$, cosmological parameters $\boldsymbol{\Omega}$, and a structure formation model $G$. The final density field $\boldsymbol{\delta}$ at redshift $z=0$ is thus related to the initial white-noise field $\mathbfit{x}$ through
\begin{equation}
    \boldsymbol{\delta} = G(\mathbfit{x},\boldsymbol{\Omega}) = G_{\mathrm{EM}} \circ G_{1\mathrm{LPT}}(\mathbfit{x},\boldsymbol{\Omega}),
\end{equation}
where we explicitly model the joint forward model $G$ as the function composition of two parts, the first being the \texttt{BORG-1LPT} model including the application of the transfer function and the primordial power spectrum, and the second being the field-level emulator. Note that the network weights and biases of the emulator are implicitly assumed.

A schematic of the incorporation of the field-level emulator into \texttt{BORG}, which we call \texttt{BORG-EM}, is shown in Fig.~\ref{fig:borg-em}. The initial white-noise field $\mathbfit{x}$ is evolved using \texttt{BORG-1LPT} to redshift $z=0$. We obtain the $3$-dimensional displacements $\boldsymbol{\Psi}_{1\mathrm{LPT}}$ using the $1$LPT predicted particle positions $\mathbfit{p}_{1\mathrm{LPT}}$ and the initial grid positions $\mathbfit{q}$. The displacements are corrected through the use of the emulator, yielding the updated displacements $\boldsymbol{\Psi}_{\mathrm{EM}}$ and, in turn, the particle positions through the use of Eq. \eqref{eq:final_pos}. The Cloud-In-Cell (CIC) algorithm is applied as the particle mesh assignment scheme, which gives us the effective number of particles per voxel and, subsequently, the final overdensity field $\boldsymbol{\delta}$ at $z=0$. After a likelihood computation, the adjoint gradient is back-propagated through the combined structure formation model $G$ to the initial white-noise field $\mathbfit{x}$.

As described in detail in \citet{Jasche2012}, new samples $\mathbfit{x}$ from the posterior distribution $\pi\left(\mathbfit{x} | \mathbfit{d}\right)$ can be obtained by following the Hamiltonian dynamics in the high-dimensional space of initial conditions. 
This requires computing the Hamiltonian forces, which can be obtained by differentiating the Hamiltonian potential $\mathcal{H}(\mathbfit{x}) = \mathcal{H}_{\mathrm{prior}}(\mathbfit{x}) + \mathcal{H}_{\mathrm{likelihood}}(\mathbfit{x})$. The introduction of the emulator only affects the likelihood part of the Hamiltonian, so we only need to differentiate
\begin{equation}
    \mathcal{H}_{\mathrm{likelihood}}(\mathbfit{x}) = -\ln \pi(\mathbfit{d}|G(\mathbfit{x},\boldsymbol{\Omega}))
\end{equation}
with respect to the initial conditions $\mathbfit{x}$. The chain rule yields
\begin{equation}
    \frac{\partial \mathcal{H}_{\mathrm{likelihood}}(\mathbfit{x})}{\partial \mathbfit{x}} = \frac{-\partial \ln \pi(\mathbfit{d}|\boldsymbol{\delta})}{\partial \boldsymbol{\delta}} \frac{\partial \boldsymbol{\delta}}{\partial \boldsymbol{\Psi}_{\mathrm{EM}}}\frac{\partial \boldsymbol{\Psi}_{\mathrm{EM}}}{\partial \boldsymbol{\Psi}_{1\mathrm{LPT}}}\frac{\partial \boldsymbol{\Psi}_{1\mathrm{LPT}}}{\partial \mathbfit{x}},
\end{equation}
where the new component is the matrix containing the gradients of the emulator output with respect to its input
\begin{equation}
    \frac{\partial \boldsymbol{\Psi}_{\mathrm{EM}}}{\partial \boldsymbol{\Psi}_{1\mathrm{LPT}}} = \frac{\partial G_{\mathrm{EM}}(\boldsymbol{\Psi}_{1\mathrm{LPT}})}{\partial \boldsymbol{\Psi}_{1\mathrm{LPT}}}
\end{equation}
which is accessible through auto-differentiation in \texttt{PyTorch}.  

\section{modelling Non-linear Dark Matter fields}
\label{sec:3}
Having the field-level emulator integrated into \texttt{BORG}, we now set up the data model necessary to infer the initial white-noise field from a non-linear dark matter distribution generated by an $N$-body simulation. To generate data with noise, we utilize a Gaussian data model such that Gaussian noise is added to the simulation output. We also introduce a novel multi-scale likelihood to the \texttt{BORG} algorithm, which allows balancing between 
the statistical information at small scales with our physical understanding at large scales, where the physics model performs best.

\subsection{Gaussian data model}
The final dark-matter particle positions from the $z=0$ snapshot of an $N$-body simulation are passed through the CIC algorithm to obtain the dark matter over-density field $\boldsymbol{\delta}^{\mathrm{sim}}$. The data $\mathbfit{d}$ is generated by
\begin{equation}
    \mathbfit{d} = \boldsymbol{\delta}^{\mathrm{sim}}+ \mathbf{n},
    \label{eq:data_model}
\end{equation}
where $\mathbf{n}$ is noise drawn from a zero mean Gaussian with diagonal covariance matrix $\mathbf{C}=\mathds{1}\sigma^2$.

During inference, the predicted final density field $\boldsymbol{\delta}$ from \texttt{BORG} and the data $\mathbfit{d}$ is compared with a likelihood function, from which the adjoint gradient backpropagates to the white noise field. To describe the data model in Eq. \eqref{eq:data_model}, we introduce the real-space voxel-based Gaussian likelihood 
\begin{equation}
    \ln \pi(\mathbfit{d}|\boldsymbol{\delta}) =
    -\frac{1}{2} \sum_i \left(\frac{\mathbfit{d}_i-\boldsymbol{\delta}_i}{\sigma}\right)^2,
    \label{eq:vox_likelihood}
\end{equation}
where $i$ runs over all voxels in the fields.

\subsection{Multi-scale likelihood model}

The likelihood in Eq. \eqref{eq:vox_likelihood} is voxel-wise, resulting in giving all the weight of the inference to the small scales. Our physics simulator has the largest discrepancy at those non-linear scales. To inform the data model about where we think the physics model is performing best, we introduce a multi-scale likelihood. Instead of giving all weight to the small scales, we re-balance in an information-theoretically correct way as shown below. Importantly, we only destroy and never introduce information, resulting in a conservative inference.

To balance the statistical information at small scales with our physical understanding at large scales, we embrace a partitioned approach to the likelihood function by incorporating $L$ factors, each dedicated to refining the prediction and data at specific scales before conducting a comparison. We start by decomposing the likelihood
\begin{equation}
    \pi (\mathbfit{d}|\boldsymbol{\delta})  = \prod_{l=0}^{L-1} \pi (\mathbfit{d}|\boldsymbol{\delta})^{w_l},
    \label{eq:partition_lh}
\end{equation}
which is a statistically valid approach as long as the weight factors $w_l$ satisfy the condition:
\begin{equation}
    \sum_{l=0}^{L-1} w_l = 1.
\end{equation}
We next introduce $\mathbf{K}^l$ to denote the operation of averaging the field values over neighbourhood sub-volumes spanning $k_l \equiv 2^{3l}$ voxels, where the subscript $l$ denotes the level of granularity. We ensure that the averaging process occurs over different sub-volumes at each likelihood evaluation. For the Gaussian likelihood in Eq.~\eqref{eq:vox_likelihood}, Eq.~\eqref{eq:partition_lh} becomes
\begin{equation}
    \ln \pi (\mathbfit{d}|\boldsymbol{\delta}) = -\frac{1}{2}\sum_{i,l} \left(\frac{[\mathbf{K}^l\mathbfit{d}]_i-[\mathbf{K}^l\boldsymbol{\delta}]_i}{\sigma_l}\right)^2 w_l,
    \label{eq:multi}
\end{equation}
where $\sigma_l = \sigma k_l^{-1/2}$ decreases with $l$ to account for increasingly coarser resolutions (see Appendix~\ref{app:multi_lh_variance}) and $i$ runs over all voxels in the fields at level $l$. 

With the use of the average pooling operators $\{\mathbf{K}^l\}$, some information is lost due to the smoothing process. As shown in Appendix~\ref{app:multi_lh_entropy} the difference in information entropy $H$ of the data vector $\mathbfit{d}$ between the voxel-based likelihood and the multi-scale likelihood becomes
\begin{equation}
    H(\mathbfit{d}) - \sum_l H(\mathbf{K}^l\mathbfit{d}) > 0.
\end{equation}
This means that the multi-scale likelihood results in not using all available information in the data, which comes with the advantage of implicitly establishing a framework for conducting robust Bayesian inference \citep[see more in e.g.,][]{Miller2015, Jasche2019} in the presence of potential model misspecification. 

More specifically, this approach can operate iteratively, with weight factors $w_l$ dynamically adjusted throughout the inference process. This strategy allows to initially use most of the large-scale information in the data before progressively incorporating finer details to capture more complex, non-linear scales (see section~\ref{sec:weight_scheduling} for further details). While this reminds of annealing-based methods in e.g. \citet{Modi2018, Porqueres2020, Bayer2023b}, it is worth stressing that our multi-scale likelihood approach is conceptually different through an information-theoretically rigorous decomposition in Eq~\eqref{eq:partition_lh} and the introduction of the average pooling operators.

\section{Demonstration of Inference with 
\texttt{BORG-EM}}
\begin{figure*}
    \centering
    \includegraphics{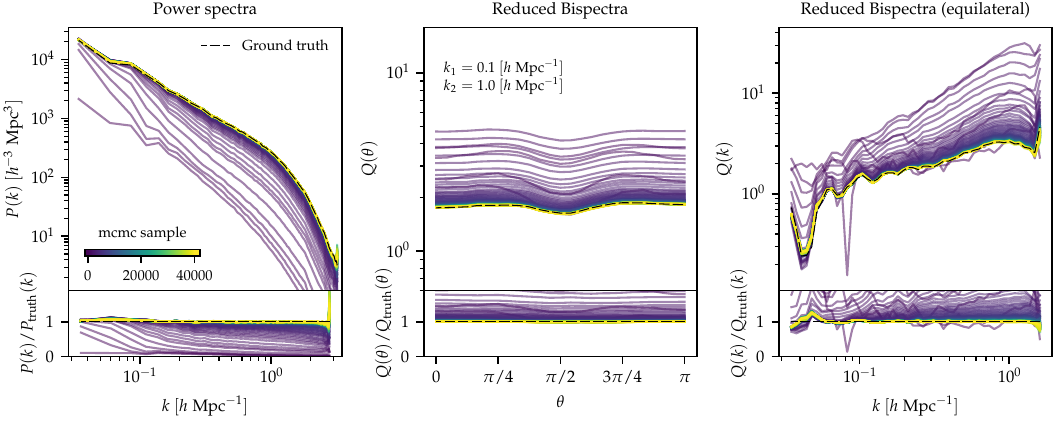} 
    \caption{The initial phase of the inference is monitored by following the systematic drift of the power spectrum and two configurations of the reduced bispectrum towards the fiducial ground truths. After a few thousand samples, we approach the regime of percent-level agreement with the power spectrum and bispectrum. More samples are needed to enter the typical set from which plausible Gaussian initial conditions can be sampled. Model misspecification between the ground truth ($N$-body) and the \texttt{BORG-EM} predicted fields shows up as small discrepancies at the largest scales, as a consequence of putting most weight on fitting the smallest scales where most statistical power resides.}
    \label{fig:warmup}
\end{figure*}

\label{sec:4}
The \texttt{BORG} algorithm uses a Markov chain Monte Carlo (MCMC) algorithm to explore the posterior distribution of white-noise fields. To evaluate the performance of the field-level emulator integration, we aim to infer the $128^3\sim10^6$ primordial white-noise amplitudes from a non-linear matter distribution as simulated by an $N$-body simulation. We use a cubic Cartesian box of side length $250h^{-1}$ Mpc on $128^3$ equidistant grid nodes, that is with a resolution of $\sim$$1.95h^{-1}$ Mpc. Notably, the emulator then operates on periodically padded displacement fields of size $224^3$ as input. To assess the algorithm's validity and efficiency in exploring the very high-dimensional parameter space, we employ warm-up testing of the Markov chain from a region far away from the target. Despite subjecting the neural network emulator to random inputs from such a Markov chain, which have not been included in the training data, it demonstrates a coherent warm-up toward the correct target region. The faster, but slightly less accurate mini-emulator is used during the initial phase of inference, after which we switch to the fully tested emulator.

\subsection{Generation of ground truth data}
\label{sec:groundtruth}
Although the forward model used during inference is limited to \texttt{BORG-EM}, that is \texttt{BORG-1LPT} plus the emulator extension, we generate the ground truth dark-matter overdensity field using the $N$-body simulation code \texttt{P-Gadget-III}, a non-public extension to the \texttt{Gadget-II} code \citep{Springel2005a}. The initial conditions are generated at $z=127$ using the IC-generator \texttt{MUSIC} \citep{Hahn2011} together with transfer file from \texttt{CAMB} \citep{Lewis1999}. 
The cosmological model used corresponds to the fiducial Quijote suite: $\Omega_\mathrm{m} = 0.3175$, $\Omega_\mathrm{b} = 0.049$, $h = 0.6711$, $n_\mathrm{s} = 0.9624$, $\sigma_8 = 0.834$, and $w=-1$ \citep{Villaescusa-Navarro2020}, consistent with latest constraints by Planck \citep{PlanckCollaboration2018}. For consistency, we compile and run the TreePM code \texttt{P-Gadget-III} with the same configuration as used for Quijote LH. This means approximating gravity consistently in terms of softening length, parameters for the gravity PM grid (grid size, $a_{\mathrm{smooth}}$), and opening angle for the tree. For example, the Quijote LH used PMGRID $=1024$, i.e. with a grid cell size of $\sim$ $0.98h^{-1}$ Mpc, which we match by using PMGRID $=256$ in our smaller volume. 

In our test scenario, we use a sufficiently high signal-to-noise ratio ($\sigma = 3$ in Eq. \eqref{eq:data_model}), such that the algorithm would have to recover all intrinsic details of the non-linearities in the large-scale structure. In contrast, a lower signal-to-noise scenario would be an easier problem as the predictions would be allowed to fluctuate more around the truth. Moreover, $\sigma = 3$ corresponds to a signal-to-noise above one for scales down to $k \approx 1.5h^{-1}$ Mpc (see Appendix~\ref{app:signal-to-noise}), which is below the scales for which the emulator is expected to perform accurately.

\subsection{Weight scheduling of multi-scale likelihood}
\label{sec:weight_scheduling}
We set up the use of the multi-scale likelihood at $L=6$ different levels of granularity, effectively smoothing the fields at most to $4^3$ voxels. The data model is then informed about the accuracy of the physics simulator at different scales. We empirically select the weights to follow a specific scheduling strategy and initialize with $w = [0.0001,0.0002,0.001,0.01,0.1,0.8887]$, i.e. with the largest weight on the largest scale. This effectively initiates the inference process at a coarser resolution to facilitate the enhancement of large-scale correlations. The weight of $w_5$ is then increased, at the expense of $w_6$, until it becomes the dominant factor. All weight transitions are shown in Appendix~\ref{app:weight_schedule}. This iterative process of increasing the weight on the next smaller scale continues until we converge after roughly $5000$ samples to empirically determined weights $\mathbfit{w} = [0.957, 0.01, 0.01, 0.009, 0.008, 0.007]$. While the majority of the weight is assigned to the smallest scale at the voxel level, non-zero weights are maintained for the larger scales. Only the last set of weights are kept when recording samples from the Markov chain, while the others belong to the warm-up phase and are discarded. While we see no major change by switching to $[1,0,0,0,0,0]$ after warm-up, further refinement of the weights is left for future research.

\subsection{Initial warm-up phase of the Markov chain}
To evaluate its initial behavior, we initiate the Markov chain with an over-dispersed Gaussian density field, set at three-tenths of the expected amplitude in a standard $\Lambda$CDM scenario. In the $128^3 \sim 10^6$ high-dimensional parameter space of white noise field amplitudes, this means that the Markov chain will initially explore regions far away from the target distribution. If it successfully transitions to the typical set and performs accurate exploration, it signifies a successful demonstration of its efficacy in traversing high-dimensional parameter space. To prematurely place the Markov chain in the target region to hasten the warm-up phase would risk bypassing a true evaluation of the algorithm's performance and exploration capabilities. 

We monitor and illustrate this systematic drift through parameter space by following the power spectrum $P(k)$ and two different configurations of the reduced bispectrum $Q$ (defined in Appendix \ref{app:powspec_and_bispec}) as computed from subsequent predicted final overdensity fields, as shown in Fig.~\ref{fig:warmup}. In particular, the middle panel shows the bispectrum as a function of the angle $\theta$ between two wave vectors, chosen to $k_1=0.1h$ Mpc$^{-1}$ and $k_2=1.0h$ Mpc$^{-1}$. In the right panel, the reduced bispectrum is displayed as a function of the magnitude of the three wave vectors for an equilateral configuration. We observe that approaching the fiducial statistics requires only a few thousand Markov chain transitions while fine-tuning to obtain the correct statistics requires additional tens of thousands of transitions during the final warm-up phase.

There are apparent differences between the resulting statistics from the forward simulated samples $\boldsymbol{\delta}$ of initial conditions. Although the small scales appear to agree with the ground truth simulation $\boldsymbol{\delta}^{\mathrm{sim}}$, there is an evident discrepancy at larger scales. The observed phenomena can be explained by the model mismatch between the field-level emulator \texttt{BORG-EM} and the $N$-body simulation. Because the multi-scale likelihood still puts the most weight on the small scales, it is expected that for the data model to fit the smallest scales, the discrepancy is pushed to the larger scales. It is worth noting that the discrepancy would be even larger for e.g. \texttt{BORG-1LPT} or \texttt{COLA}, where the model mismatch is more significant.

\section{Accuracy of inferred initial conditions}
\label{sec:5}
After the completion of the warm-up phase, the Markov chain generates samples $\{\mathbfit{x}\}$ from the posterior distribution $\pi(\mathbfit{x}|\mathbfit{d})$. We show that the inferred initial conditions display Gaussianity with the expected mean and variance, and that the transfer function with the ground truth initial conditions show below $5\%$ agreement over all scales. We also show that the cross-correlation between the final density fields and the data is as high as the correlation between the simulated $N$-body and the data, indicating we extracted all cross-correlation information content from the data. Notably, information on linear scales in the initial conditions is mainly tied up on non-linear scales in the final conditions due to gravitational collapse. This highlights the importance of non-linear physics models, such as the field-level emulator, to constrain linear regimes in the initial conditions.

\begin{figure}
    \centering
    \includegraphics{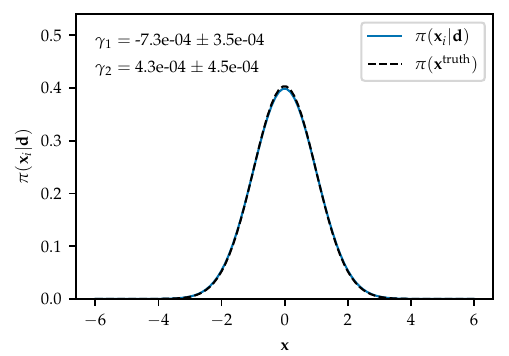}
    \caption{The distribution of individual samples of initial conditions $\{\mathbfit{x}_i\}$, here displayed next to the true initial conditions $\mathbfit{x}_{\mathrm{true}}$ for data generation, exhibit desired Gaussian properties: zero mean, unit variance, and minimal skewness $\gamma_1$ and excess kurtosis $\gamma_2$ (with standard deviation errors computed over the samples) \vspace{0.5em}.}
    \label{fig:post-dist}
    \includegraphics{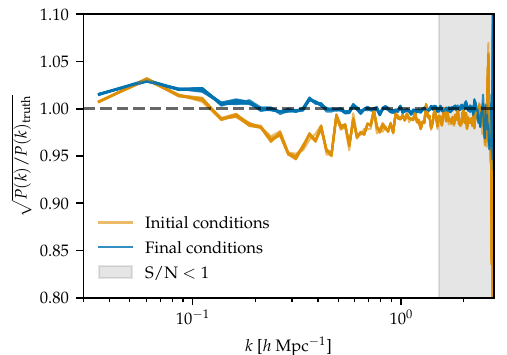}
    \caption{Transfer functions for the inferred initial and final density fields with the respective ground truth show agreements within $5\%$ across the entire Fourier range. The noise-dominated region of the data is shown in grey. Notably, the final conditions align more closely with the ground truth than the initial conditions. We expect this and other percent-level deviations to result from small-scale model misspecifications, which propagate through mode coupling and power conservation to larger scales in the initial conditions.}
    \label{fig:transfer_ic}
\end{figure}

\begin{figure*}
    \centering
    \includegraphics{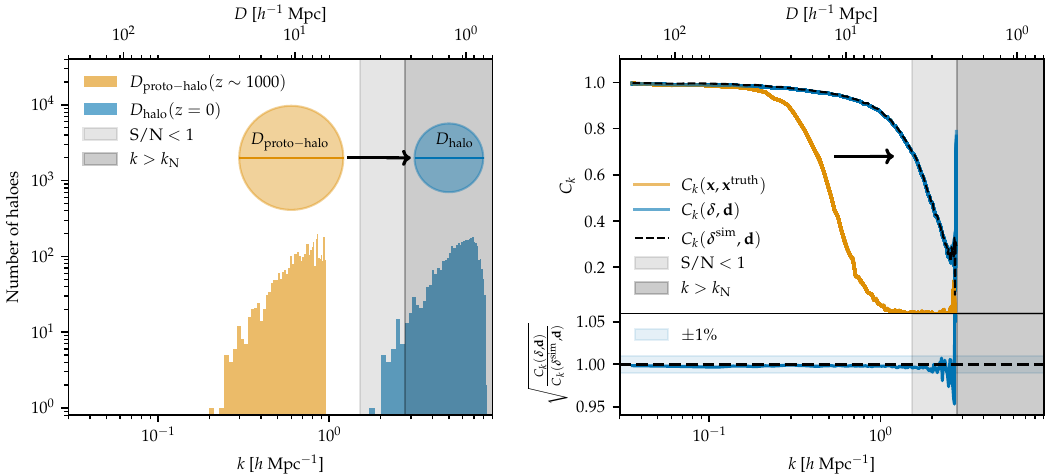}
    \caption{Information recovery of inferred initial conditions. Left: the size of proto-haloes (as defined by the diameter $D_{\mathrm{proto-halo}}$) in the initial conditions collapse into smaller haloes (of size $D_{\mathrm{halo}}$) in the final field. In the spherical collapse approximation to halo formation, all of our proto-haloes collapse into the noise-dominated region (S/N $<1$) of our data, and most of them even below our data grid resolution as given by the Nyquist frequency $k_\mathrm{N}$. Right: Cross-correlation analysis reveals that the inferred initial fields $\mathbfit{x}$ exhibit notable correlation with the truth $\mathbfit{x}^{\mathrm{truth}}$ up to $k \sim 0.35h$ Mpc$^{-1}$. Between the corresponding final conditions $\boldsymbol{\delta}$ and the data $\mathbfit{d}$, this correlation extends to much smaller scales of $k \sim 2h$ Mpc$^{-1}$. Note that this closely mirrors the correlation between the true final density field $\boldsymbol{\delta}^{\mathrm{sim}}$ and the data $\mathbfit{d}$, which demonstrates that we have fully extracted the cross-correlation information, as evident in the lower panel. Importantly, this is obtained despite the lower correlation in the initial field. With our noise level and data grid resolution, and because larger regions in the initial conditions collapse into smaller regions in the final conditions, there is no more information in the initial conditions to gain.}
    \label{fig:corr}
\end{figure*}

\subsection{Statistical accuracy}
To assess the quality of the drawn initial condition samples from the posterior distribution, we first verify their expected Gaussianity as compared to the ground truth in Fig.~\ref{fig:post-dist}, including tests of skewness and excess kurtosis going to zero as expected. Notably, in Fig.~\ref{fig:warmup}, we also see that the two- and three-point statistics (power spectra and bispectra) of the inferred initial conditions towards the end of warm-up are well-recovered to below a few percent and $10\%$ respectively. 

We also evaluate the transfer function, defined as the square root of the ratio between the power spectra of the inferred fields and the ground truth. As illustrated in Fig.~\ref{fig:transfer_ic}, we achieve high accuracy ($<5 \%$) for both the initial and final conditions across all modes. We observe that the final conditions align more closely with the ground truth. This is the result of accurately explaining the $N$-body generated data with the emulator during inference despite the residual model misspecification, particularly small-scale inaccuracies (see e.g. Fig.~\ref{fig:fwd_pow_bi}). The percent-level deviations in the transfer functions at intermediate and large scales can also be attributed to model misspecifications, which propagate through mode coupling (as discussed along Fig.~\ref{fig:corr} in section~\ref{sec:info_recovery}) and power conservation in cosmological fields to affect larger scales in the inferred initial conditions. Furthermore, as most statistical information resides at smaller scales, the emulator model prioritizes fitting these scales, further explaining the minor larger-scale discrepancies. The quality of the initial conditions remains high, as shown in section~\ref{sec:6}.

\subsection{Information recovery}
\label{sec:info_recovery}
From the collapsed objects in the final conditions, we are trying to recover the information on the initial conditions. In this pursuit we have to acknowledge the effects of gravitational collapse, causing objects that are initially extended objects in Lagrangian space to collapse to smaller scales in the final conditions. This means that the gravitational structure formation introduces a temporal mode coupling between the present-day small scales and the earlier large scales, as illustrated in Fig.~\ref{fig:corr}.

\subsubsection{Proto-haloes vs haloes}
We distinguish between the Lagrangian scales in the initial conditions and the Eulerian scales in the final conditions due to mass transport. In particular, overdense regions in the initial conditions collapse into smaller regions in the final conditions, while underdense regions experience growth \citep{Gunn1972}. We can understand this mode coupling in more detail by spherical collapse approximation to halo formation. We start by considering a halo of mass $M$ and follow the trajectories of all of its constituent particles back to the initial conditions. Based on the early Universe being arbitrarily close to uniform with a mean density $\rho_\mathrm{m} = \rho_\mathrm{c} \Omega_\mathrm{m}$,
we can imagine the collapse of a uniform spherical volume – the \textit{proto-halo} volume – from which the present-day halo formed. We can define the proto-halo through the \textit{Lagrangian radius} $R_L$ within which all particles from a halo with mass $M$ must have initially resided,
\begin{equation}
     R_L = \left(\frac{3M}{4\pi\rho_m}\right)^{1/3}.
\end{equation}
We thus expect the linear regime in the initial conditions to be pushed into the nonlinear regime of today. 

Using the $M_{200\mathrm{c}}$ mass definition (see more in section~\ref{sec:haloes}) we find haloes in the ground truth simulation with accurate mass estimates in the range $[0.853,16] \times 10^{14} M_{\odot}$, corresponding to Lagrangian radii between $6.14h^{-1} \; \mathrm{Mpc}$ and $16.3h^{-1} \;\mathrm{Mpc}$ for the least and most massive haloes respectively. The diameter $D=2R_L$ can be converted to the corresponding scale of these regions $k \in [0.193, 0.512]h \; \mathrm{Mpc}^{-1}$. Note, however, that in Fig.~\ref{fig:corr} we show all haloes found, even the smallest ones. The virial radii for these haloes lie in the range $[0.34, 1.85]h^{-1} \; \mathrm{Mpc}$, i.e. with diameter scales of $k \in [1.70, 9.24]h \; \mathrm{Mpc}^{-1}$. In the left panel of Fig.~\ref{fig:corr} we highlight this size difference of haloes and the corresponding proto-haloes in units of the corresponding scale in $h \; \mathrm{Mpc}^{-1}$. As our Cartesian grid of the final density field has a resolution of $1.95h^{-1}$ Mpc with a Nyquist frequency $k_{\mathrm{N}} = 2.79h \; \mathrm{Mpc}^{-1}$, this means that the majority of proto-haloes will collapse to haloes below the resolution of our data constraints. While one might anticipate an impact on the reconstructed halo centers, our observations suggest the effect is minor, as presented in section~\ref{ssec:Stacked halo density profiles}. The precise extent to which additional information can be extracted at higher resolution remains uncertain and requires further investigation.

\subsubsection{Limitations set by Gaussian noise and data grid resolution}
We also note that because of non-linear structure formation, different cosmic environments in the data will be differently informative at different scales \citep[see e.g.][]{Bonnaire2022}. This is particularly evident in our case of using a Gaussian data model since Gaussian noise, in contrast to e.g. Poisson noise, is insensitive to the environment of the matter distribution. With our fixed noise threshold, the signal-to-noise ratio becomes significantly higher in over-dense regions as compared to under-dense regions. It is therefore expected that the constraining power in our inference will come almost exclusively from overdense regions.

In Fig.~\ref{fig:corr} we show where the transition from signal domination to noise domination occurs for our use of $\sigma=3$ in Eq. \eqref{eq:data_model} (also see Appendix~\ref{app:signal-to-noise}). This introduces a soft barrier below which recovery of information is limited. We also display a hard boundary of information recovery given by the data grid resolution, corresponding to the three-dimensional Nyquist frequency $k_{\mathrm{N}} \approx 2.79h \; \mathrm{Mpc}^{-1}$, below which no information is accessible.

\subsubsection{Cross-correlation in inferred fields}
\label{sec:cross}
We examine the information recovered from our inference by computing the cross-correlation between the true white noise field $\mathbfit{x}^{\mathrm{truth}}$ and the set of sampled fields $\{\mathbfit{x}\}$, as depicted in the right panel of Fig.~\ref{fig:corr}. We see a strong correlation ($>$$80\%$) with the truth up to $k \sim 0.35h$ Mpc$^{-1}$, after which the correlation drops. We also compare the correlation between the final density fields $\boldsymbol{\delta}$ and the data $\mathbfit{d}$. Furthermore, we benchmark this to the correlation between the true final density field $\boldsymbol{\delta}^{\mathrm{sim}}$ and the data $\mathbfit{d}$. As depicted in the lower panel, we achieve a correlation for the inferred fields that is nearly as high, differing by only a sub-percent margin up to $k \sim 2h$ Mpc$^{-1}$. This demonstrates that the cross-correlation information content between the final fields $\boldsymbol{\delta}$ and the data $\mathbfit{d}$ has been fully extracted well within the noise-dominated regime up to the grid resolution of the data. 

Despite achieving maximal information extraction from the final conditions, we are limited in the reconstruction of initial conditions by noise and data grid resolution. This is promising since it shows that more information can be obtained. For instance, the use of a lower signal-to-noise ratio exhibit diminished correlation as shown in Appendix~\ref{app:signal-to-noise-2}, indicating that higher signal-to-noise ratios increase information extraction. Because most of our probes moved into these regimes (high noise or even sub-grid), by enhancing the resolution in the data space we could potentially also boost recovery and increase correlation in the initial conditions up to smaller scales. We leave this for future investigation, but it is clear that to constrain the linear regime of the initial conditions, nonlinear models are essential.

\section{Posterior Resimulations}
\label{sec:6}
As highlighted by \citet{Stopyra2023} in the context of model misspecification, a sufficiently accurate physics model is needed during inference to recover the initial conditions accurately. Otherwise, when using the inferred initial conditions to run posterior resimulations in high-fidelity simulations, one may risk obtaining an overabundance of massive haloes and voids as observed by \citet{Desmond2021,Hutt2022,Mcalpine2022}. To validate the incorporation of the emulator in field-level inference, we use $80$ independent posterior samples (see details in Appendix \ref{app:acf}) of inferred initial conditions within the $N$-body simulator \texttt{P-Gadget-III}, mirroring the data generation setup (see section~\ref{sec:groundtruth}). The posterior resimulations align with the ground truth $N$-body simulation in the formation of massive large-scale structures, in terms of halo mass function as well as density profiles, which demonstrates the robustness of the emulator model in inference. The mean and standard deviation of the posterior resimulations as well as of the inferred initial conditions are shown in Fig.~\ref{fig:post_slices}.

\subsection{Halo mass function}
\label{sec:haloes}
\begin{figure*}
    \centering
\includegraphics{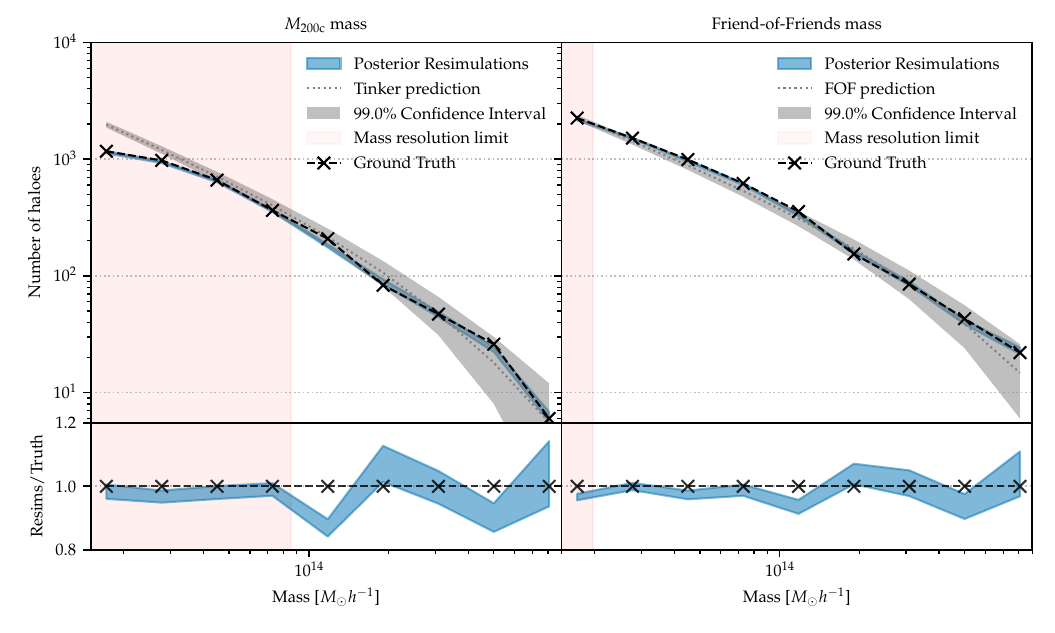}
    \caption{Halo mass function for the entire cubic volume of side length $250h^{-1}$ Mpc obtained using posterior resimulations of the \texttt{BORG-EM} field-level inference. For comparison with the ground truth simulation, we adopt two different halo mass definitions, $M_{200\mathrm{c}}$ (through \texttt{AHF}) (left) and \texttt{FoF} (right). Note that $M_{200\mathrm{c}}$, being a spherical overdensity definition, puts stricter requirements on the density profile. The posterior resimulation bins are plotted along with uncertainties only originating from the scatter in the resimulations to emphasize its minimal size, demonstrating our ability to accurately determine the count in each bin. To show that our resimulations are consistent with the halo mass function (i.e., within the Poisson interval of the expected mean in that bin) we plot the theoretical halo mass function with Poisson scatter. The grey shaded regions show the $99\%$ confidence interval from Poisson fluctuations around the Tinker and \texttt{FoF} predictions, respectively. The mass function is not accurately reproduced below the mass resolution limit, illustrated by the red shaded region and defined as haloes consisting of less than $130$ and $30$ simulated particles for $M_{200\mathrm{c}}$ and \texttt{FoF}, respectively.}
    \label{fig:pHMF}
\end{figure*}

To identify haloes in our $N$-body simulations we employ the spherical-overdensity based halo finder \texttt{AHF} \citep{Knollmann2009} and the Friend-of-Friends (\texttt{FoF}) algorithm in \texttt{nbodykit} \citep{Hand2017a}. We adopt the $M_{200\mathrm{c}}$ definition (for \texttt{AHF}) for the halo mass, representing the mass enclosed within a spherical halo whose average density is equal to $200$ times the critical density. Recovering the $M_{200\mathrm{c}}$ mass of haloes is generally a more stringent test than the \texttt{FoF} technique because it requires resolving the density profile accurately down to smaller scales. \texttt{FoF} does not depend on a fixed density threshold to define halo boundaries. Instead, it determines boundaries based on particle distribution and connectivity using a linking length (we use $l = 0.2h^{-1}$ Mpc). This allows for more forgiving interpretations of halo size and shape, not requiring precise density profiles for larger regions. Consequently, \texttt{FoF} haloes demonstrate more robustness against density variations and are adept at accommodating complex substructures.

In Fig.~\ref{fig:pHMF}, we show the halo mass function as found by \texttt{AHF} and \texttt{FoF}, clearly demonstrating the high accuracy of the posterior resimulations. The number density of haloes from the posterior resimulations shows high consistency both with \texttt{FoF} and $M_{200\mathrm{c}}$ haloes from the ground truth simulation. We also validate the $M_{200\mathrm{c}}$ halo mass function against the theoretically expected \texttt{Tinker} prediction for this cosmology \citep{Tinker} (represented by the grey shading). Similarly, the \texttt{FoF} halo mass function is validated against theory prediction \citep{Watson2012}. We conclude that the resimulations, plotted with the scatter across the resimulation samples, align with the predicted halo mass function, i.e., fall within the Poisson interval of the expected mean. While also possible to include the Poisson error for the resimulations into a single error, the Poisson error would dominate, obscuring an important point: the resimulation scatter is minimal, allowing for precise determination of the count in each bin. This is a result of the high signal-to-noise ratio employed in this study, along with the fact that all voxels are constrained by data. The residual panel further reveals small discrepancies, with some bins containing more halos and others fewer in the posterior resimulations compared to the ground truth. These discrepancies can be attributed to model misspecification between the emulator and the $N$-body ground truth during inference.

A recent investigation on the number of particles required to accurately describe the density profile and mass of $M_{200\mathrm{c}}$ dark matter haloes has shown that at least $N=130$ particles are required \citep{Mansfield2020}. This results in the Poisson contribution $\sqrt{N}/N$ to the halo mass being below $10\%$, ensuring greater accuracy. With our simulation set-up, this corresponds to $8.53\times 10^{13} M_{\odot}$, below which the mass is in general not accurately recovered. The (red) shaded region of Fig.~\ref{fig:pHMF} represents the mass resolution limit. In our case, masses down to $5\times 10^{13} M_{\odot}$ show a significant correlation with the theoretical expectation. Below the mass resolution limit, we also match the ground truth mass function down to $2\times 10^{13} M_{\odot}$. The \texttt{FoF} mass resolution limit is set to $30$ particles (in our case corresponding to $2\times 10^{13} M_{\odot}$) which is sufficient to accurately recover the mass function \citep{Jenkins2001,Trenti2010}, as shown here next to the theoretical prediction from \citet{Watson2012}.

\subsection{Stacked halo density profiles}
\label{ssec:Stacked halo density profiles}
\begin{figure*}
    \centering
    \includegraphics{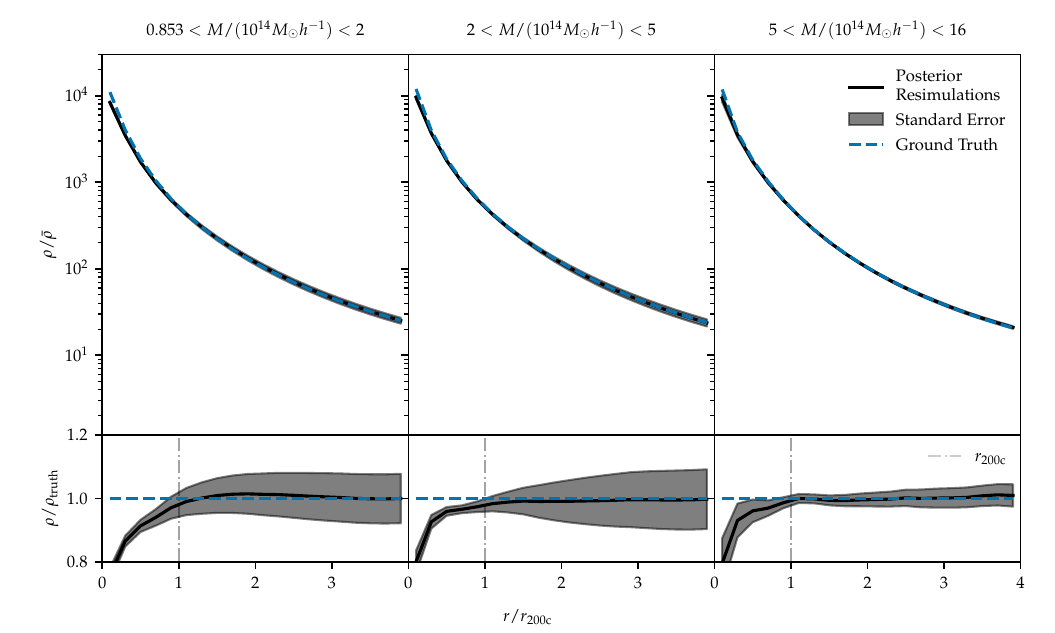}
    \caption{Stacked halo density profiles obtained from the posterior resimulations and the ground truth in three distinct mass bins. We used the ground truth $N$-body simulation prediction as an initial guess for the locations of each halo in each posterior resimulation and re-centered these with an iterative scheme to identify the correct centre of mass for each halo. The average density profile within each mass bin is then computed for each resimulation, after which we average the density profiles over all resimulations. In all bins, we demonstrate discrepancies below $10\%$ for $r>\frac{1}{2}r_{200\mathrm{c}}$. Note that the virial radius $r_{200\mathrm{c}}$ for the ground truth haloes lie in the range $[0.34,1.85]h^{-1}$ Mpc, i.e. below the data resolution.}
    \label{fig:post_prof}
\end{figure*}

In addition to the halo mass functions in Fig.~\ref{fig:pHMF}, which shows high consistency with $\Lambda$CDM expectations and the ground truth, we investigate how well the density profiles of haloes in different mass bins are reconstructed. The centre of mass of each halo will, however, vary from one posterior sample to the next. The mean displacement of halo centres over all posterior resimulations is found to be $0.13^{+0.03}_{-0.02}h^{-1}$ Mpc. We, therefore, use the ground truth $N$-body simulation prediction as an initial guess for the locations of each halo in each posterior resimulation and re-center these with an iterative scheme to identify the correct centre of mass for each. This prescription allows us to identify the haloes in posterior resimulations corresponding to each of the ground truth haloes, which in turn enables us to compare how well we constrain the halo density profiles. Notably, this results in only using the haloes present in the ground truth simulation. 

We use the cumulative density profile  
\begin{equation}
    \rho(r) = \frac{M(<r)}{4\pi r^3/3},
\end{equation}
and group haloes into three different mass bins. We compute the mean density profile for each mass bin in each posterior resimulation, and subsequently average the means over the simulations. 

For the lower mass bin, we pick the $8.53\times 10^{13} M_{\odot}$ threshold as described in the previous section, and for the high mass end, we select $1.6\times 10^{15} M_{\odot}h^{-1}$ as it includes all haloes in the simulation. In units of $10^{14}M_{\odot}$, the bins are $[0.853,2]$, $[2,5]$ and $[5,16]$. 

We display the average over the stacked density profiles as well as the standard error in Fig.~\ref{fig:post_prof}. For all mass bins, as compared to the ground truth, we see at most a $10\%$ discrepancy down to $\frac{1}{2}r_{200\mathrm{c}}$, highlighting the high quality of the sampled initial conditions by \texttt{BORG-EM}.

\section{Discussion \& Conclusions}
\label{sec:7}
The maximization of scientific yield from upcoming next-generation surveys of billions of galaxies heavily relies on the synergy between high-precision data, high-fidelity physics models, and advanced technologies for extracting cosmological information. In this work, we built on recent success in analysis at the field level, which naturally incorporates all non-linear and higher-order statistics of the cosmic matter distribution within the physics model. More specifically, we incorporated a field-level emulator into the Bayesian hierarchical inference algorithm \texttt{BORG} to infer the initial white-noise field from a dark matter-only density field produced by a high-fidelity $N$-body simulation. While reducing computational cost compared to full $N$-body simulations, the emulator significantly increases the accuracy as compared to first-order Lagrangian perturbation theory and \texttt{COLA}. As outlined by \citet{Stopyra2023}, a sufficiently accurate physics model is required to accurately reconstruct massive objects, a standard that our emulator surpasses.

In this work, we introduced a novel multi-scale likelihood approach that balances our comprehension of large-scale physics with the statistical power on smaller scales. This approach effectively informs the data model about the physics simulator's accuracy across various scales during inference. The inevitable loss of data information as shown by information entropy due to the average operator introduces a method for conducting robust Bayesian inference, proving advantages in instances of model misspecification.

In our demonstration, we utilized the dark-matter-only output of the $N$-body simulation code \texttt{P-Gadget-III} added with Gaussian noise as the data. Within the Hamiltonian Monte Carlo based \texttt{BORG} algorithm, we sampled the initial conditions in the form of $128^3 \sim 10^6$ primordial white-noise amplitudes in a cubic volume with side length $250h^{-1}$ Mpc and with a grid resolution of $1.95h^{-1}$ Mpc. 

A worry in putting a trained machine-learning algorithm into an MCMC algorithm is that the MCMC algorithm might probe regions, particularly during the initial phase, that have not been part of the training domain of the network. Therefore, the MCMC framework could be trapped in an error regime of the machine-learning algorithm. Our observations suggest that this never occurs and that the emulator as well as the mini-emulator thus show that they have become sufficiently generalized during training to allow for exploration within an MCMC framework. This could partly be explained by $1$LPT playing a pivotal role in fixing the global configuration, while the emulator is a fairly local and sparse operator that only updates the smaller scales.

Having reached the desired target space in the $\sim $$10^6$ high-dimensional space of initial conditions, we note that the sampled initial conditions display the expected Gaussianity. The percent-level agreement for the power spectrum and bispectrum compared to the ground truth reveals our ability to accurately recover scales down to approximately $k\sim2h$ Mpc$^{-1}$. 

Importantly, upon examining information recovery, we note that the inference has fully extracted all cross-correlation information that can be gained from the noisy data up to the data grid resolution. This is evident by the sub-percent level agreements as compared to the correlation between the ground truth simulation and the data. The initial conditions are constrained up to $k \sim 0.35h$ Mpc$^{-1}$ at the $80\%$ level, demonstrating that non-linear data models are essential to constrain even the linear regime of the initial conditions. Spherical collapse theory, in which massive objects encompasses larger scales in the initial conditions, offers one explanation. The Eulerian scale of the haloes has virialization radii that are below our data resolution. This is promising as it indicates that increasing the resolution of the data could potentially increase the recovered small-scale information in the initial conditions. We leave the investigation of increasing the data resolution for future work. 

We further demonstrate the robustness of incorporating the field-level emulator in inference by running posterior resimulations, using the inferred initial conditions. We note consistency for the stringent $M_{200\mathrm{c}}$ definition of halo mass as well as for the Friend-of-friends halo mass function down to $0.853\times 10^{14}M_{\odot}$. We further show less than $10\%$ discrepancies for stacked density profiles in all mass bins

The current emulator network boasts improved accuracy and speed, yet there's a chance the architecture isn't fully optimized. Particularly, our mini-emulator in this study reveals a promising trend: despite substantially reduced neural network model weights, leading to a considerable cut in evaluation time, the accuracy remains notably high. This finding merits further investigation. 
The potential of field-level emulators shown in this work further suggests that an extension to more complicated scenarios such as using hydrodynamical simulations might be possible. Complex physics can be directly integrated into the training data using the same pipeline showcased in this work. Of interest to this work would also be extending the emulator to multiple redshifts. Such a redshift-dependent emulator has already been developed by \cite{Jamieson2024}. Redshift is added as a second style parameter, and the emulator is retrained on a suite of simulations including snapshots that span a range of redshifts from $z=3$ to $z=0$. Another redshift-dependent emulator has also been developed by \cite{Bartlett2024} using the scale factor as a style parameter.

Given that the emulator operates as an extension to the structure formation model module and is trained on multiple cosmologies, there is also interest in integrating this technique with the exploration of cosmological parameters. Notably, joint sampling of the initial density field and cosmological parameters within the \texttt{BORG} framework has been investigated in prior research \citep{Ramanah2018,Andrews2022,Porqueres2023}.

In conclusion, we here demonstrated that Bayesian inference within a Hamiltonian Monte Carlo machinery with a high-fidelity simulation is numerically feasible. Machine learning in the form of field-level emulators for cosmological simulation makes it possible to extract more of the non-linear information in the data.
\section*{Acknowledgements}
We thank Adam Andrews, Metin Ata, Nai Boonkongkird, Simon Ding, Mikica Kocic, Stuart McAlpine, Hiranya Peiris, Andrew Pontzen, Natalia Porqueres, Fabian Schmidt, and Eleni Tsaprazi for useful discussions related to this work. We also thank Adam Andrews and Nhat-Minh Nguyen for their feedback on the manuscript. This research utilized the Sunrise HPC facility supported by the Technical Division at the Department of Physics, Stockholm University. This work was carried out within the Aquila Consortium\footnote{\href{https://www.aquila-consortium.org/}{https://www.aquila-consortium.org/}}. This work was supported by the Simons Collaboration on “Learning the Universe”. This work has been enabled by support from the research project grant ‘Understanding the Dynamic Universe’ funded by the Knut and Alice Wallenberg Foundation under Dnr KAW 2018.0067. This work has received funding from the Centre National d’Etudes Spatiales. LD acknowledges travel funding supplied by Elisabeth och Herman Rhodins minne. SS is supported by the Göran Gustafsson Foundation for Research in Natural Sciences and Medicine. JJ acknowledges support by the Swedish Research Council (VR) under the project 2020-05143 -- ``Deciphering the Dynamics of Cosmic Structure". JJ acknowledges the hospitality of the Aspen Center for Physics, which is supported by National Science Foundation grant PHY-1607611. The participation of JJ at the Aspen Center for Physics was supported by the Simons Foundation.

\section*{Data Availability}
The data underlying this article will be shared on reasonable request to the corresponding author.
\endgroup

\bibliographystyle{mnras}
\bibliography{ref} 

\appendix

\section{Field-level emulator}
We here describe in more detail the neural network architecture used, the generation of consistent training data with the first-order LPT implementation in \texttt{BORG}, and the changes necessary to train a mini-emulator.

\subsection{Neural network architecture}
\label{app:NN}
We implement a U-Net/V-Net architecture, akin to the one in \citet{Jamieson2022} and \citet{AlvesDeOliveira}. This model operates on four resolution levels in a "V" pattern, utilizing a series of $3$ downsampling (by stride-$2$ $2^3$ convolutions) and subsequently $3$ upsampling layers (by stride-$1/2$ $2^3$ transposed convolutions). Each level involves $3^3$ convolutions, with the incorporation of a residual connection (a $1^3$ convolution instead of identity) within each block, inspired by ResNet \citep{He2016}. Batch normalization layers follow every convolution, except the initial and final two, and are accompanied by a leaky ReLU activation function (with a negative slope of $0.01$). Similar to the U-Net structure, inputs from downsampling layers are concatenated with outputs from the corresponding resolution levels' upsampling layers. Each layer maintains $64$ channels, excluding the input and output layers (3) and those following concatenations (128). A difference from the original U-Net is that we directly add input displacement to the output.

We use the same loss function as in \citet{Jamieson2022b},
\begin{equation}
    \log L = \log L_{\delta} + \lambda \log L_{\boldsymbol{\Psi}},
    \label{eq:NN_loss}
\end{equation}
where $L_{\delta}$ and $L_{\boldsymbol{\Psi}}$ are mean square error losses on the Eulerian density field $\delta$ and particle displacements $\boldsymbol{\Psi}$ respectively, and the weight parameter is chosen to $\lambda=3$ for faster training.

\subsection{Generation of \texttt{BORG-LPT} data}
\label{app:borglpt}
The emulator from \citet{Jamieson2022b} was retrained with input data coming from the implementation of $1$LPT in \texttt{BORG}. \texttt{BORG-1LPT} was run down to $z=0$. The initial conditions used for Quijote \citep{Villaescusa-Navarro2020} have a mesh size of $1024^3$, which sets the highest resolution that can be run. The initial conditions generator computes the $2$LPT displacements at $z=127$, and fills the mesh up to the particle grid Nyquist mode, and uses zero padding on the rest, as this effectively de-aliases the convolutions when computing the $2$LPT corrections \citep[see e.g.][]{Michaux2021}. 

For the Latin hyper-cube, the particle grid used for the $N$-body simulation was $512^3$. In preparation for using the white noise field in \texttt{BORG}, we make use of the idea behind the zero padding by setting all modes higher than the particle grid Nyquist mode to zero. We then run \texttt{BORG-LPT} with $1024^3$ particles and finally downsample by removing every other particle in each dimension.

\subsection{Mini-emulator}
\label{app:distill}
We downsize the neural network architecture presented in Appendix \ref{app:NN} by altering the channel count in each layer from $64$ to $32$, consequently reducing the number of model weights from $3.35 \times 10^6$ to $0.84 \times 10^6$.

To reduce the computational cost for training, we only make use of $5\%$ of the full data set. Based on ideas from recent knowledge distillation techniques \citep{Wang}, we also introduce a novel approach of leveraging the outputs of the emulator to train the mini-emulator. In essence, the comprehensive emulator acts as an auxiliary teacher, complementing the Quijote data. The loss  $L$ from Eq. \eqref{eq:NN_loss} thus becomes only a part in the new $L_{\mathrm{mini}}$ loss through
\begin{equation}
    \log L_{\mathrm{mini}} = \frac{1}{2}\log L + \frac{1}{2}L_{\mathrm{teacher}},
\end{equation}
where the individual losses $L$ and $L_{\mathrm{teacher}}$ both have the form of Eq. \eqref{eq:NN_loss}, with the only difference being the ground truth output either as the Quijote suite or the full emulator predictions. This student-teacher methodology yields a modest yet discernible enhancement over the performance of a teacher-less mini-emulator  
(i.e., when training the mini-emulator independently), but further investigation is needed.

\section{Power Spectrum and Bispectrum}
\label{app:powspec_and_bispec}
The density modes, denoted as $\delta(\mathbfit{k})$, are obtained by performing the Fourier transform on the density field $\delta(\mathbfit{s})$ at each location $\mathbfit{s}$ in space. The power spectrum $P(k)$, which describes the distribution of the density modes, can then be computed as:
\begin{equation}
    \left\langle\delta(\mathbfit{k}) \delta\left(\mathbfit{k}^{\prime}\right)\right\rangle=(2 \pi)^3 \delta_{\mathrm{D}}^3\left(\mathbfit{k}+\mathbfit{k}^{\prime}\right) P(k) 
\end{equation}
In this expression, $\delta_{\mathrm{D}}^{(3)}(\mathbfit{k})$ denotes the 3D Dirac delta function, which enforces the homogeneity of the density fluctuations. The requirement of isotropy is reflected in the fact that the power spectrum only depends on the magnitude $k \equiv|\mathbfit{k}|$. Similarly, the cross-power $C_k$ spectrum between two fields $\delta_1$ and $\delta_2$ is given by
\begin{equation}
    \left\langle \delta_1(\mathbfit{k}) \delta_2(\mathbfit{k}^{\prime}) \right\rangle = (2\pi)^3 \delta_{\mathrm{D}}^3(\mathbfit{k} + \mathbfit{k}^{\prime}) C_k(k),
\end{equation}
where we usually normalize $C_k$ with the product $P_1(k)P_2(k)$.

The three-point statistic in the form of the matter bispectrum $B\left(k_{1:3}\right)$, where $k_{1:3} \equiv \{k_1, k_2, k_3\}$, is given by
\begin{equation}
\left\langle\delta(\mathbfit{k}_1)\delta(\mathbfit{k}_2) \delta(\mathbfit{k}_3)\right\rangle= (2 \pi)^3 \delta_{\mathrm{D}}^{(3)}\left(\sum_{i=1}^3\mathbfit{k}_i\right) \times B\left(k_{1:3}\right).
\end{equation}
Note that the sum of the wave vectors in the Dirac delta function requires them to form a closed triangle. The bispectrum can therefore be expressed through the magnitude of two wave vectors and the separation angle $\theta$, i.e. $B\left(k_1, k_2, \theta \right)$. We also define the reduced bispectrum $Q$ as
\begin{equation}
    Q\left(k_1, k_2, k_3\right) \equiv \frac{B\left(k_1, k_2, k_3\right)}{P_1 P_2+P_2 P_3+P_3 P_1},
\end{equation}
which has a weaker dependence on cosmology and scale. To compute the power spectrum, cross-power spectrum, and bispectrum, we make use of the \texttt{PYLIANS3}\footnote{\href{https://github.com/franciscovillaescusa/Pylians3}{https://github.com/franciscovillaescusa/Pylians3}} package \citep{Pylians}. 

\section{Approximate physics models}
\label{app:fwd_models}
\begin{figure}
    \centering
\includegraphics{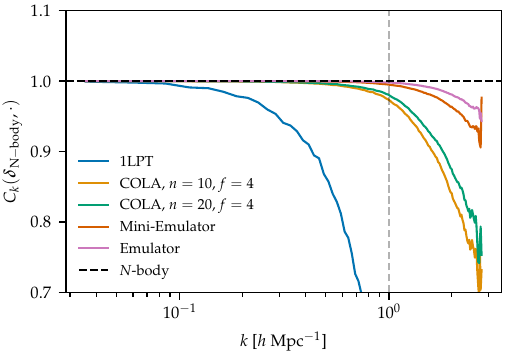}
    \caption{Comparison between different gravity models in terms of the cross-power spectrum. Both the mini-emulator and emulator outperform first-order LPT and \texttt{COLA} (using $n=10$ and $n=20$ time-steps; force factor $f=4$) and reach a percent-level agreement with the $N$-body simulation predictions.}
    \label{fig:fwd_cross}
\end{figure}

\begin{figure*}
    \centering
\includegraphics{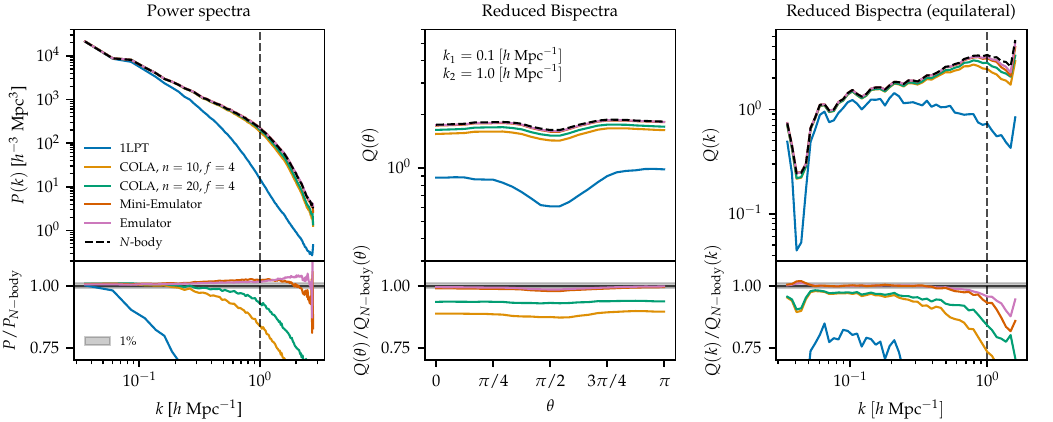}
    \caption{Comparison between different gravity models, all having been run with $128^3$ particles in a cubic volume with side length $250h^{-1}$ Mpc box to $z=0$ using the same initial conditions, in terms of the power spectrum and two configurations of the reduced bispectrum. Both the mini-emulator and emulator outperform first-order LPT and \texttt{COLA} (using $n=10$ and $n=20$ time-steps; force factor $f=4$) and reach a percent-level agreement with the $N$-body simulation predictions.}
    \label{fig:fwd_pow_bi}
\end{figure*}

\begin{figure*}
    \centering \includegraphics{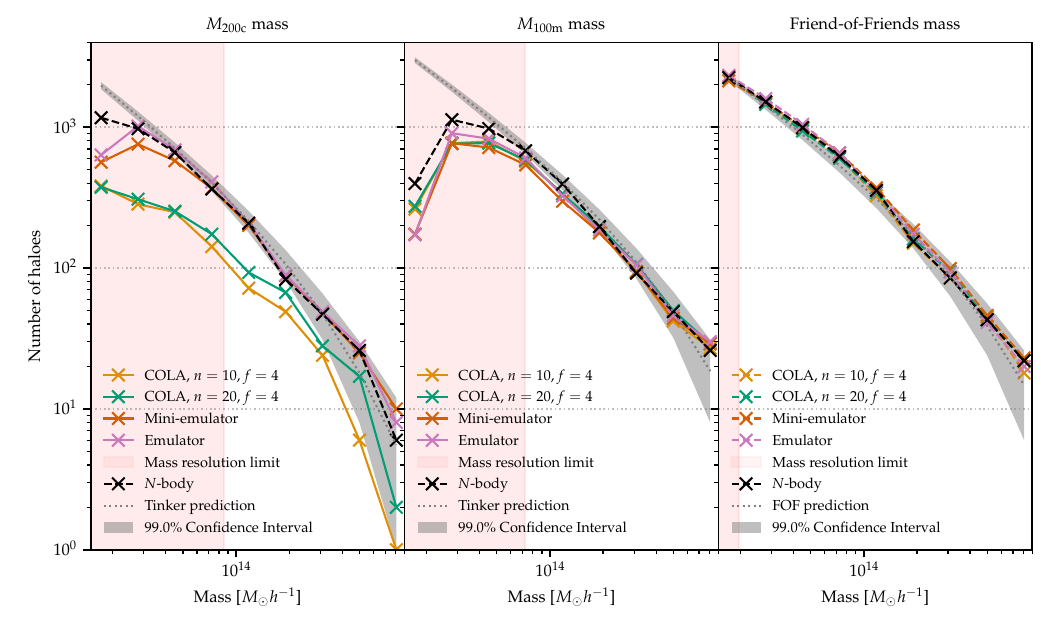}
    \caption{Halo mass function for different gravity models, all having been run with $128^3$ particles in a cubic volume with side length $250h^{-1}$ Mpc box to $z=0$ using the same initial conditions, computed using three different halo mass definitions. Left: $M_{200\mathrm{c}}$ (mass is the total mass within a radius such that the average density is $200$ times the critical density of the Universe). Middle: $M_{100\mathrm{m}}$ (mass within a sphere such that the radius is $100$ times the mean density of the Universe). We validate the $M_{100\mathrm{m}}$ and $M_{200\mathrm{c}}$ halo mass functions against the theoretically expected \texttt{Tinker} prediction for this cosmology \citep{Tinker}. Right: \texttt{FoF} mass with linking length $0.2$, as compared with the mass function from \citet{Watson2012}. Notably, the inability of LPT to resolve small scales results in the absence of halos that can be reliably identified across all defined halo definitions. Below the mass resolution limit, which differs between the mass definitions ($130$, $110$, and $30$ particles, respectively) as given by \citet{Jenkins2001,Trenti2010,Mansfield2020}, we cannot accurately recover the mass function. As compared to \texttt{COLA} (using $n=10$ and $n=20$ time-steps; force factor $f=4$) the emulators display significant improvement towards the $N$-body mass functions.}
    \label{fig:fwd_HMF}
\end{figure*}

\begin{figure*}
    \centering
    \includegraphics{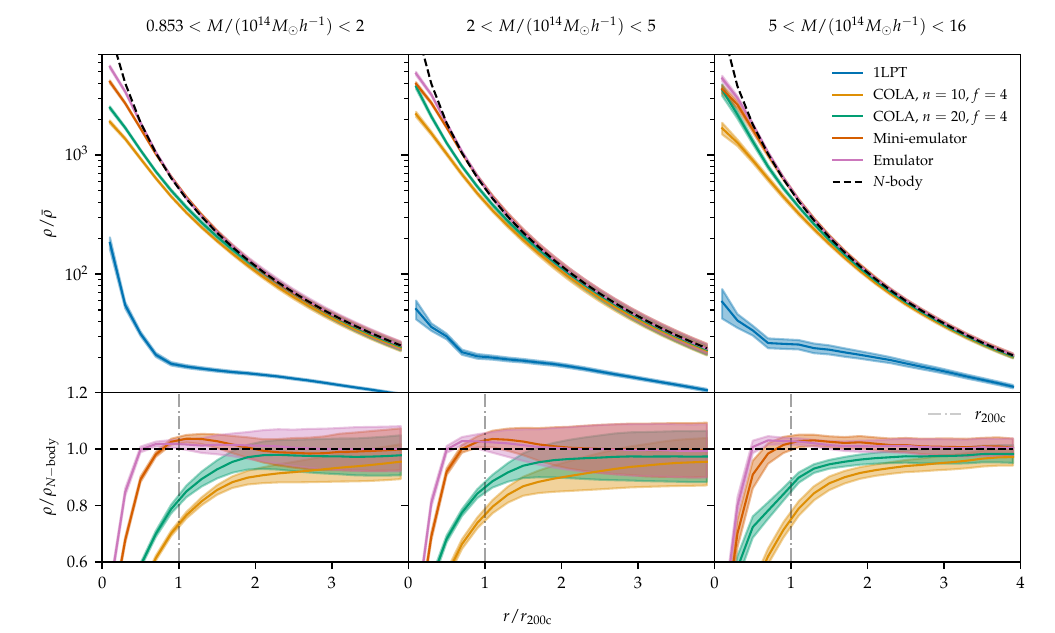}
    \caption{Stacked density profiles of haloes for different gravity models using the same initial conditions with $128^3$ particles in a cubic volume with side length $250h^{-1}$ Mpc. To identify halo locations we used the $N$-body predictions and re-centred them to determine the centre of mass for each object in every forward model. This allows us to define halo density profiles for first-order LPT even though we cannot reliably identify halos. We adopt the $M_{200\mathrm{c}}$ mass definition and average the profile over all haloes in three mass bins. Only the emulators show accurate profiles below the virial radius $r_{200\mathrm{c}}$ as compared to the $N$-body. Note that the virial radius $r_{200\mathrm{c}}$ for the ground truth haloes lie in the range $[0.34,1.85]h^{-1}$ Mpc. }
    \label{fig:fwd_density}
\end{figure*}

We compare the different structure formation models \texttt{BORG-$1$LPT}, \texttt{COLA} (using force factor $f=4$ and either $n=10$ or $n=20$ time steps), mini-emulator, and emulator with the high-fidelity $N$-body simulation code \texttt{P-Gadget-III} (see section~\ref{sec:groundtruth}). This comparison encompasses an examination of the cross power spectrum in Fig.~\ref{fig:fwd_cross} and the power spectrum and two distinct configurations of the reduced bispectrum in Fig.~\ref{fig:fwd_pow_bi} (see Appendix~\ref{app:powspec_and_bispec}). Moreover, three different definitions of halo mass function in Fig.~\ref{fig:fwd_HMF}, and density profiles in Fig.~\ref{fig:fwd_density}. In all statistics, it is evident that the emulator, as well as the mini-emulator, outperforms the other approximate model and provides highly accurate predictions compared to the $N$-body predictions. We used the $N$-body simulation prediction to identify halo locations and re-centred these to identify the centre of masses of the same objects in each forward model.

Of most relevance is the emulators' improved accuracy over \texttt{COLA}. In terms of the power spectrum and the reduced bispectra, the emulators improve from a $>$$10\%$ discrepancy at small scales down to percent-level discrepancies up to $k \sim 2h^{-1}$ Mpc for the fiducial cosmology, as can be seen in Fig.~\ref{fig:fwd_pow_bi}. 

Furthermore, while \texttt{COLA} performs accurately for $M_{100\mathrm{m}}$ and the Friend-of-Friends mass functions (Fig.~\ref{fig:fwd_HMF}), \texttt{COLA} struggles significantly with the $M_{200\mathrm{c}}$ mass function as well as the inner halo density profiles 
 (Fig.~\ref{fig:fwd_HMF} and Fig.~\ref{fig:fwd_density}). This follows from the fact that obtaining accurate $M_{200\mathrm{c}}$ masses requires accurately determining the density profile at the $200\rho_c$ radius, which corresponds to about $\sim630$ times the mean density (of the Universe) for our value of $\Omega_\mathrm{m}$. This occurs at the radius $r_{200\mathrm{c}}$, and in Fig.~\ref{fig:fwd_density} this is where \texttt{COLA} is inaccurate, while the emulators are still highly accurate. Hence, the emulators are able to get the correct $M_{200\mathrm{c}}$ masses down to the mass resolution limit. $M_{100\mathrm{m}}$ and Friend-of-Friends mass functions, however, are less strict and do not require density profiles to be accurate on scales as small as the $r_{200\mathrm{c}}$. This explains why \texttt{COLA} performs almost as accurately as the emulators on these mass functions. 

\section{Multi-scale likelihood}
We here derive the effect on the variance of averaging the fields and the effect this in turn has on the information entropy of the data.  We also provide the weights used for different scales during inference.

\subsection{Scale-dependent variance}
\label{app:multi_lh_variance}
\noindent
At the target resolution $N = 128^3$, the noise is drawn from a zero-mean Gaussian with covariance matrix $\mathbf{C}$, such that noise realizations  $\mathbf{n}\curvearrowleft \pi(\mathbf{n})=\mathcal{N}(\mathbf{0},\mathbf{C})$ represent independent Gaussian noise corresponding to the Gaussian data model in Eq.~\eqref{eq:data_model}. We assume a constant data variance \(\sigma^2\) for all diagonal elements of the noise covariance matrix \(\mathbf{C}\), whose elements thus are
\begin{equation}
    \mathbf{C}_{ij} = \int n_i n_j \pi(\mathbf{n}) \mathrm{d}\mathbf{n} = \langle n_i n_j \rangle = \delta^K_{ij} \langle n_i^2 \rangle = \delta^K_{ij} \sigma^2
\end{equation}
where $n_i$ and $n_j$ represent field values of the noise at voxels $i$ and $j$, and the Kronecker delta $\delta^K_{ij}$ describes the diagonal covariance matrix.

We now apply the average pooling operator $K^l$ defined as
\begin{equation}
    K_{ir}^l = \begin{cases}
        \; k_l^{-1} \quad & \mathrm{if} \quad i\in \Omega_r \\
        \; 0 & \mathrm{otherwise}
    \end{cases}
\end{equation}
where $i$ and $r$ represent voxels in the original and smoothed field, respectively, and $\Omega_r$ is the domain of $k_l = 2^{3l}$ adjacent voxels from the original field that are averaged over to level $l$. The total number of voxels in the smoothed grid becomes $N_l = N k_l^{-1}$. The operation can now be defined through
\begin{equation}
    n_r^l = \sum_i K_{ir}^ln_i,
\end{equation}
where $n_r^l$ is the field value of the noise at voxel $r$ at smoothing level $l$. To derive the variance at the smoothing level $l$, we evaluate the covariance matrix at this level as
\begin{align}
    \mathbf{C}_{rs}^l & = \int (\mathbf{K}^l \mathbf{n})_r (\mathbf{K}^l \mathbf{n})_s \pi(\mathbf{n}) \mathrm{d}\mathbf{n} \nonumber \\
    & =  \int \left( \sum_i K_{ir}^l n_i \right) \left( \sum_j K_{js}^l n_j \right) \pi(\mathbf{n}) \mathrm{d}\mathbf{n} \nonumber \\
    & = \sum_{i,j} K_{ir}^l K_{js}^l \int n_i n_j \pi(\mathbf{n}) \mathrm{d}\mathbf{n} = \sum_{i,j} K_{ir}^l K_{js}^l \delta^K_{ij} \langle n_i^2 \rangle \nonumber \\
    & = \sum_i K_{ir}^l K_{is}^l \sigma^2 = \delta^K_{rs} \sigma^2 \sum_i \left(K_{ir}^l\right)^2 \nonumber \\
    & = \delta^K_{rs} \sigma^2 |\Omega_r| \left(\frac{1}{k_l}\right)^{2}  = \delta^K_{rs} \frac{\sigma^2}{k_l},
\end{align}
where, in the ultimate step, we utilized that there are $k_l$ voxels in the domain $\Omega_r$ of the smoothing process. Additionally, we accounted for the non-overlapping nature of the smoothing domains $\Omega_r$ and $\Omega_s$, leading to the product of the kernels \(K_{ir}^l\) and \(K_{is}^l\) being non-zero only if \(r = s\). The resulting relation
\begin{equation}
    \sigma_l^2 = \frac{\sigma^2}{k_l}
    \label{eq:variance_level_l}
\end{equation}
was here derived for $\mathbf{n}$ but note it also holds for $\mathbf{d}=\boldsymbol{\delta}^{\mathrm{sim}}+\mathbf{n}$ since $\boldsymbol{\delta}^{\mathrm{sim}}$ is fixed and the variance only enters through $\mathbf{n}$. \\

\subsection{Information entropy}
\label{app:multi_lh_entropy}
The information entropy $H$ for the data field $\mathbfit{d}$ is defined as
\begin{equation}
    H(\mathbfit{d}) = -\int p(\mathbfit{d})\log p(\mathbfit{d}) \mathrm{d}\mathbfit{d} = -\mathbb{E}[\log p(\mathbfit{d})]
\end{equation}
We identify $p(\mathbfit{d}) \equiv \pi(\mathbfit{d}|\boldsymbol{\delta})$ as our likelihood and perform the standard calculation of entropy of an independent multi-variate Gaussian, yielding
\begin{equation}
  H(\mathbfit{d}) = \frac{N}{2} \left(1+\log(2\pi) + 2 \log\sigma\right),
  \label{eq:entropy_voxel}
\end{equation}
where we used that the covariance matrix in our case is diagonal with variances $\sigma^2$ and $N=128^3$ is the total number of voxels. We proceed similarly to the sum in Eq~\eqref{eq:multi}: after applying a smoothing operator $\mathbf{K}^l$ separately, we compute the information entropy and then sum the individual information entropies. Similar to Eq~\eqref{eq:entropy_voxel} and using our result from Eq~\eqref{eq:variance_level_l}, we obtain

\begin{align}
    H(\mathbf{K}^l\mathbfit{d}) & =\frac{N_lw_l}{2} \left(1+\log(2\pi) + 2\log\sigma_l \right)  \\ & =  \frac{Nw_l}{2k_l} \left(1+\log(2\pi) +2 \log\left(\frac{\sigma}{k_l}\right) \right).
    \label{eq:entropy_multi}
\end{align}

The total loss of information can be written
\begin{equation}
    H(\mathbfit{d}) - \sum_l H(\mathbf{K}^l\mathbfit{d}) > 0,
\end{equation}
which is evident since $k_l=2^{3l}>1$, and $\sum_l w_l = 1$, making each of the three terms in Eq.~\ref{eq:entropy_multi} lower than in Eq.~\ref{eq:entropy_voxel}, even when summed over all $l$. This indicates that the multi-scale likelihood erased information in the data $\mathbfit{d}$.

\subsection{Weight scheduling}
\label{app:weight_schedule}
In Table~\ref{tab:weights} we show the weight values used in the multi-scale likelihood during the initial phase of the inference. Importantly, weights are only adjusted during the warm-up phase. The final set of weights, corresponding to the last row of the table, is used when recording samples.

\begin{table}
\centering
\begin{tabular}{|c|c|c|c|c|c|c|}
\toprule
$N_{\textrm{accept}}$ & $w_1$ & $w_2$ & $w_3$ & $w_4$ & $w_5$ & $w_6$ \\
\midrule
0 & 0.0001 & 0.0002 & 0.001 & 0.01 & 0.1 & 0.8887 \\
300 & 0.0001 & 0.0002 & 0.001 & 0.1 & 0.8887 & 0.01 \\
700 & 0.0001 & 0.0002 & 0.1 & 0.8797 & 0.01 & 0.01 \\
1200 & 0.0001 & 0.1 & 0.8699 & 0.01 & 0.01 & 0.01 \\
1800 & 0.1 & 0.86 & 0.01 & 0.01 & 0.01 & 0.01 \\
2500 & 0.957 & 0.01 & 0.009 & 0.009 & 0.008 & 0.007 \\ 
\midrule
\midrule
\end{tabular}
\caption{The six-stage weight schedule within the multi-scale likelihood during the initial phase of the inference. The weights are provided corresponding to specific numbers of accepted samples $N_{\textrm{accept}}$ in the Markov chain. A linear increase is applied between the specified weight values. The values at the transitions were chosen empirically to ensure that the scale with the highest $w$ made the most significant likelihood contribution.}
\label{tab:weights}
\end{table}

\section{Signal-to-noise}
\label{app:signal-to-noise}
\begin{figure}
    \centering
    \includegraphics{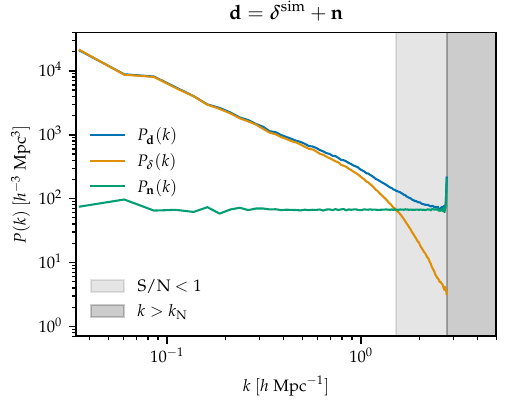}
    \caption{Power spectra for the ground truth simulation $\boldsymbol{\delta}^{\mathrm{sim}}$, the added noise $\mathbf{n} \curvearrowleft \pi(\mathbf{n})=\mathcal{N}(\mathbf{0},\mathds{1}\sigma^2)$ with $\sigma=3$, and the data $\mathbf{d}$. The grey regions show where the signal-to-noise is greater than $1$ (for $k>1.52h$ Mpc$^{-1}$) and the Nyquist frequency $k_{\mathrm{N}} \approx 2.79h$ Mpc$^{-1}$.}
    \label{fig:noise_level}
\end{figure}

The Gaussian noise is insensitive to density amplitudes, resulting in high signal-to-noise in denser regions and low signal-to-noise in underdense regions. In Fig.~\ref{fig:noise_level} we show that for $k>1.52h$ Mpc$^{-1}$ the noise dominates over the signal. We also indicate the Nyquist frequency, corresponding to the smallest scale from which information can be accessed.

\section{Inference with lower signal-to-noise ($\sigma=4$)}
\label{app:signal-to-noise-2}
\begin{figure}
    \centering
    \includegraphics{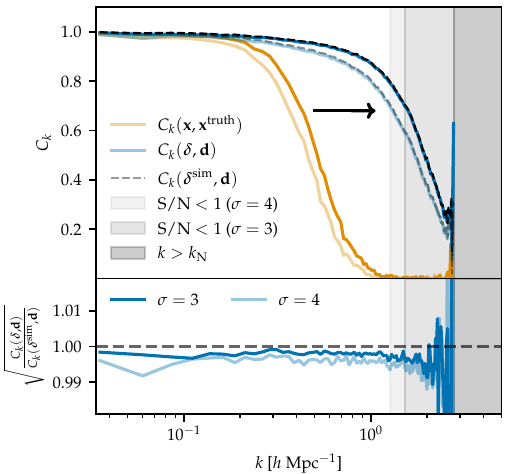}
    \caption{Information recovery of inferred initial conditions as in Fig.~\ref{fig:corr}, but here with $\sigma=4$ used in Eq.~\eqref{eq:data_model} for generating the mock data and in Eq.~\eqref{eq:multi} during inference. The results corresponding to $\sigma=3$, as shown in Fig.~\ref{fig:corr}, are represented by the bold lines, while the $\sigma=4$ case is shown by fainter lines. In both cases, the mean cross-correlation over the inferred samples is displayed. In line with expectations, this lower signal-to-noise scenario yields reduced correlation for both initial and final conditions.}
    \label{fig:corr_sigma4}
\end{figure}

The signal-to-noise chosen affects the amount of information that can be extracted. In a sibling inference with $\sigma=4$, i.e. with a lower signal-to-noise, we obtain the expected behaviour of lower correlations over the full Fourier range, as shown in Fig.~\ref{fig:corr_sigma4}.

\setcounter{figure}{0}
\renewcommand{\thefigure}{H\arabic{figure}}
\begin{figure*}
    \centering
    \includegraphics{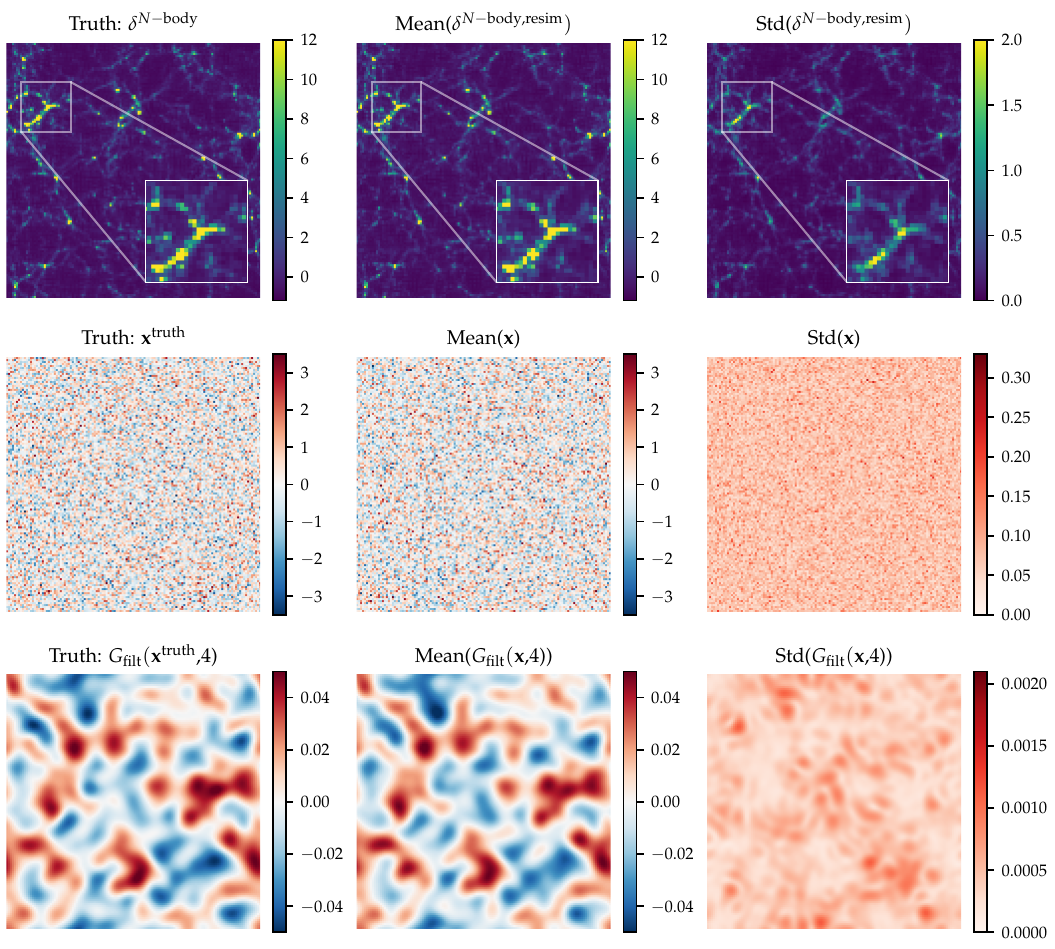}
    \caption{The mean of the posterior resimulations (top center) is compared with the ground truth (top left) using a projection slab width of $19.5h^{-1}$ Mpc. The mean exhibits a high visual correlation with the ground truth down to the smallest (voxel) scales. The standard deviation (top right) varies non-uniformly following the large-scale structure. The inferred initial conditions (bottom two rows; slab of width $1.95h^{-1}$ Mpc) show no correlation with the ground truth at the smallest scale, as also illustrated quantitatively by Fig.~\ref{fig:corr} and Fig.~\ref{fig:corr_sigma4}. However, employing a Gaussian smoothing filter, here with a smoothing scale $\sigma_{\mathrm{filt}}=4$, visually reveals the correlation with the truth (bottom row).}
    \label{fig:post_slices}
\end{figure*}

\section{Auto-correlation function}
\label{app:acf}
To check how many samples are needed to obtain independent samples from the posterior, we compute the auto-correlation function $\text{ACF}(l)$ for the white noise field $\mathbfit{x}$ as 
\begin{equation}
\text{ACF}(l) = \frac{1}{4N} \cdot \mathcal{F}^{-1}\left[\mathcal{F}(\mathbfit{x}_i - \bar{\mathbfit{x}}_i) \star (\mathcal{F}(\mathbfit{x}_i - \bar{\mathbfit{x}}_i))^*\right]_l,
\end{equation}
where $l$ here is the lag for a given Markov chain for an individual white noise amplitude $\mathbfit{x}_i$ with mean $\bar{\mathbfit{x}}_i$. Note also that $N=128^3$ is the total number of amplitudes, $\mathcal{F}$ is the Fast Fourier Transform, $\star$ denotes a conjugation and $^*$ represents a complex conjugate. 

We establish the correlation length of the Markov chain as the sample lag needed for the auto-correlation function to drop below $0.1$, and obtain an average of $430 \pm 154$ samples. 

\section{Slices of posterior fields}
\label{app:samples}
\noindent
In Fig.~\ref{fig:post_slices} we show slices of the posterior fields, both of the inferred initial conditions and the resimulations.

\bsp
\label{lastpage}

\end{document}